\documentclass[superscriptaddress,showpacs,aps,twocolumn,prb,floatfix]{revtex4}
\usepackage{bm,amsmath,amssymb}
\usepackage{graphicx,color}

\begin{document}

\title{Helical spin thermoelectrics controlled by a side-coupled magnetic quantum dot in the quantum spin Hall state}

\author{Pablo Roura-Bas}
\affiliation{Centro At\'omico Bariloche, CNEA, 8400 S. C. de
Bariloche, Argentina}

\author{Liliana Arrachea}
\affiliation{International Center for Advanced Studies, Escuela de Ciencia y Tecnolog\'{\i}a, Universidad Nacional de San Mart\'{\i}n,  25
de Mayo y Francia, 1650 Buenos Aires, Argentina}
\affiliation{Kavli Institute for Theoretical Physics, Kohn Hall, University of California at Santa Barbara, Santa Barbara, Santa Barbara, CA 93106, USA}

\author{Eduardo Fradkin}
\affiliation{Department of Physics and Institute for Condensed Matter Theory,
University of Illinois at Urbana-Champaign, 1110 West Green Street, Urbana, Illinois 61801-3080, USA}

\begin{abstract}
We study the thermoelectric response of a device containing a pair of helical edge states contacted at the same temperature $T$ and chemical potential $\mu$ and connected to an external reservoir, with different chemical potential and
temperature, through a side quantum dot. Different operational modes can be induced by applying   a magnetic field $B$ and a gate voltage $V_g$ at the quantum dot. At finite $B$, the quantum dot acts simultaneously  as a  charge and a spin filter. Charge and spin currents are induced, not only through the quantum dot, but also along the edge states. We focus on linear response and analyze the regimes, which we  identify as charge heat engines or refrigerator, spin heat engine and spin refrigerator.
\end{abstract}

\pacs{ }
\maketitle

\section{Introduction}
\label{intro}

One of most remarkable properties of topological insulating phases is the existence of conducting edge states. In two-dimensional systems, such as in the quantum Hall state or the
spin quantum Hall  (QSH) state, these edge states are single or multiple  one-dimensional (1D) channels through which charge and energy propagate only in one direction. 
In the case of the quantum Hall, the edge states are chiral \cite{wen, halperin, laughlin, buttiker, fradkin}
as a consequence of the broken time-reversal symmetry. Instead, the QSH hosts Kramers pairs of helical counter-propagating edge states with opposite spin orientation. \cite{ti1,ti2,ti3,ti5,ti6,ti7}
 In situations where the
electron-electron interactions play a role, these systems exhibit another extraordinary feature: the  fractionalization of the charge and spin in the low-energy excitations. This has an impact in
the unidirectional transport properties along the edges, \cite{wen,k-f-1,k-f-2,fradkin} as well as through tunneling contacts to other edge states or to other structures like quantum dots. \cite{chamon,eun-ah,chang,glattli,rosenow,torsten,enhan} The edge states  of the fractional quantum Hall effect are well described as chiral Luttinger liquids and the fractional charge is directly related to
the magnetic filling factor. \cite{wen,fradkin} A pair of edge states of the QSH is effectively described by a Luttinger liquid of left and right moving electrons where spin and  charge fractionalization can take place as a consequence of the Coulomb interaction represented by a parameter $K$. \cite{gia,fradkin}

Thermal transport in edge states of the quantum  Hall  regime has received a great deal of  attention for some time. Since the pioneer works by Kane y Fisher, \cite{k-f-1, k-f-2} several studies were reported on heat transport in the integer \cite{granger,Nam,us,qcond} and in the fractional regimes.
\cite{cappelli,grosfeld,us1,stern,heiblum,altimiras, yacoby, altimiras2, baner,pheno} More recently, thermoelectric effects have also been explored in the integer, \cite{rafa,peter} as well as
in the fractional case. \cite{enhan,vanuci} Relatively less is presently known about  thermal transport and thermoelectric effects in the quantum spin Hall regime. \cite{jauho,rone,roda,graph,bjorn,soth,benj}

\begin{figure}[h]
\vspace{1.cm}
\begin{center}
\includegraphics[width=\columnwidth]{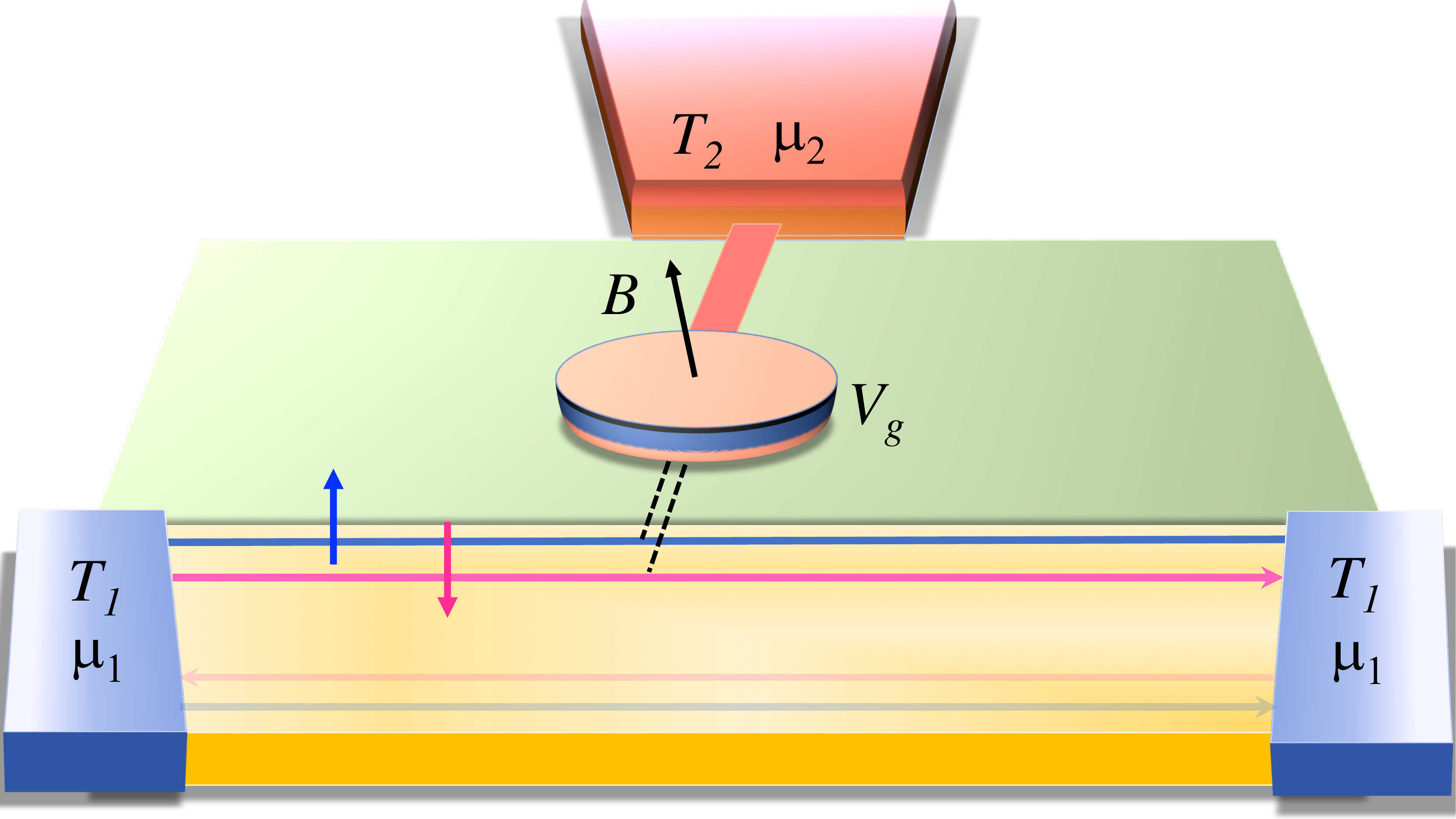}
\end{center}
\caption{Sketch of the setup. A Kramers pair of helical edge states of the spin Hall effect is ballistically contacted to terminals at temperature $T_1$ and chemical potential $\mu_1$, while
tunneled contacted to an external reservoir at a different temperature $T_2$ and chemical potential $\mu_2$ through a quantum dot. The transport 
through the quantum dot can be controlled by means of a gate voltage and a weak magnetic field $B$. }
\label{fig0}
\end{figure}

Tunneling contacts to quantum dots in mesoscopic structures play a crucial role in the thermoelectric response of these systems. This is because they enable transport with broken
particle-hole symmetry, which is a necessary condition for realizing the charge to energy conversion characteristic of thermoelectricity. \cite{linke,casati,ora,thir,tang} Charge transport in helical edges of the quantum spin Hall regime with 
tunneling contacts between edge states and quantum dots or antidots is a subject of very active theoretical investigation. \cite{dolcini,citro,bruno1,tidot1,tidot2,tidot3,tidot4,tidot5,tidot6,tidot7,tidot8,tidot9,tidot10,tidot11,tidot12,tidot13,bruno2}

The aim of the present work is to analyze the thermoelectric response of quantum spin Hall helical edges  realized  in  the structure sketched in Fig. \ref{fig0}. It consists of a pair of helical edge states of a quantum spin Hall system connected to
an external reservoir at a different temperature and chemical potential, through a tunneled-coupled quantum dot. The transport through the quantum dot is controlled by means of a gate voltage 
and a magnetic field. As we will discuss, the physics of this device is very rich, allowing for the implementation of a variety of thermoelectric regimes, taking advantage of the helical nature of the edge states. The possibility of
the magnetic control of the charge flow in helical edges was explored in other setups with magnetic islands in direct contact to the edge states. \cite{magdot1,magdot2,magdot3} 
Here
we consider the effect of a Zeeman field on the side-coupled dot of the structure of the figure, in order to introduce spin filtering of the electrons injected from the external reservoir. The magnetic field is localized at the quantum dot and it is weak enough to leave the helical edge states unaffected. With these ingredients  we have, not only heat-charge conversion, but also heat-spin conversion. 

Without the ingredient of the magnetic field at the quantum dot, the setup of the figure may operate as a charge heat engine, or refrigerator, as is usual in thermoelectric devices where heat flows induced by temperature differences can be  used to generate charge currents against chemical potential differences and vice versa. 
The magnetic field at the quantum dot adds the
possibility of inducing spin currents, not only through the quantum dot, but also through the helical channels between the left and right terminals of the QSH bar. This comes along with heat flow between these
two terminals, even when they are assumed to be at the same temperature and chemical potential. We identify two relevant operational regimes  introduced by the spin filtering, which we name the spin heat engine and  and the spin refrigerator, respectively. We analyze in detail all these regimes. On the other hand, the fractionalization of charge and energy in chiral Luttinger liquids has the consequence of enhancing the thermoelectric performance of the fractional quantum Hall regime, relative to the integer case,  
described by  non-interacting electrons. \cite{enhan} We show that the Coulomb repulsion represented by the Luttinger parameter $K$ has a similar effect in the present case. 

The paper is organized as follows. In Section II we present the theoretical model for the setup of Fig. \ref{fig0}. In Section III we present   the theoretical framework to analyze its thermoelectric response.
We present results for the different operational modes in Section IV. Summary and conclusions are presented in Section V.

\section{Model}
\label{model}
The theoretical model for the setup of Fig. \ref{fig0} is defined by the following Hamiltonian
\begin{equation}\label{H}
 H = H_{edges} + H_{dot} + H_{tun},
\end{equation}
where the first term correspond to the pair of helical edges. It is modeled as a Luttinger liquid as follows\cite{fradkin,gia}
\begin{equation}\label{Hll}
H_{edges} = \frac{v}{4\pi K}\int dx \Bigl [  \bigl( \partial_x\phi_{\uparrow}(x)\bigr)^2 + 
                                           \bigl( \partial_x\phi_{\downarrow}(x)\bigr)^2 \Bigl ].
\end{equation}
Notice that due to the helical nature of the
edge states, the direction of propagation is determined by the spin orientation. Hence, the spin labels $\uparrow$ and $\downarrow$ denote at the same time left and right moving excitations propagating with velocity $v$.
  The Coulomb electron-electron interaction is
characterized by the parameter $K$, such that $K<1$  ($K>1$) corresponds to repulsive (attractive) interactions and  $K=1$ corresponds to the non-interacting case.
  The densities $ \partial_x \phi_{\sigma}(x)$, are expressed in terms of  bosonic modes obeying a Kac Moody algebra
\begin{equation}
\left[\phi_{\downarrow}(x), \phi_{\downarrow}(x')\right] = -\left[\phi_{\uparrow}(x), \phi_{\uparrow}(x')\right] =  i \pi K \mbox{sgn}(x-x^{\prime}).
\end{equation}
They are related to  fermionic fields through
\begin{equation}\label{fermion-fields}
\psi_{\downarrow,\uparrow}(x) =\frac{F_{\downarrow,\uparrow}}{\sqrt{2\pi a}}~e^{i[K_{\pm}\phi_{\downarrow}(x)+K_{\mp}\phi_{\uparrow}(x)]}
\end{equation}
with $a$ being a characteristic length and 
\begin{equation}
K_{\pm} = (K^{-1} \pm 1)/2.
\end{equation}
The Klein factors $F_{\sigma}$ ensure $\big\{\psi_{\sigma}(x),\psi^{\dagger}_{\sigma'}(x') \big\}=\delta_{\sigma,\sigma'} \delta(x-x')$.

The second term of Eq. (\ref{H}) describes a quantum dot side-coupled to the QSH bar. It is 
controlled by a gate voltage $V_g$ and a magnetic field $B$, which has components $B_{||}$, parallel, and $B_{\perp}$, perpendicular, to
 the direction of the spin-orbit coupling of the QSH system. We focus on the situation where the magnetic field is strongly localized at the side-coupled quantum dot and weak enough to preserve the time-reversal invariance of the QSH bar. 
 For simplicity, we consider a single-level quantum dot,
\begin{equation} \label{hd}
H_{dot}=\sum_{\sigma} \varepsilon_{d \sigma} d^{\dagger}_{\sigma} d_{\sigma} + \varepsilon_{\perp} \left( d^{\dagger}_{\uparrow} d_{\downarrow} + H. c. \right),
\end{equation}
with $\varepsilon_{d, \sigma}= e V_g + s_{\sigma} \mu_B B_{||}/2$, $s_{\uparrow,\downarrow}= \pm$ and $\varepsilon_{\perp}=  \mu_B B_{\perp}/2$. 
It is convenient to diagonalize the Hamiltonian for the dot as follows
\begin{equation}
H_{dot}=\sum_{s=\pm} E_s d^{\dagger}_s d_s.
\end{equation}
The local energies in the diagonal basis read
\begin{equation}\label{eigenvalue}
%E_s=\frac{ \left( \varepsilon_{\uparrow} + \varepsilon_{\downarrow} \right)\pm r}{2},  \;\;\;\;\;\; r = \sqrt{\left( \varepsilon_{\uparrow} - \varepsilon_{\downarrow} \right)^2 + 4 \varepsilon_{\perp}^2}.
E_s=e V_g \pm \mu_B \frac{B}{2},  \;\;\;\;\;\; B = \sqrt{B^2_{||} + B_{\perp}^2}.
\end{equation}
The change of basis is $d_s= \sum_{\sigma} u_{s,\sigma} d_{\sigma}$, with
\begin{eqnarray}\label{change}
 u_{+,\uparrow}&=& \cos(\theta/2),\;\;\;u_{-,\uparrow}= \sqrt{1-u_{+,\uparrow}^2}\nonumber\\
 u_{+,\downarrow}&=& -u_{-,\uparrow},\;\;\;u_{-,\downarrow}=u_{+,\uparrow},
\end{eqnarray}
where the angle $\theta$ is determined by $\cos(\theta)=B_{||}/B$.

The quantum dot is tunneled coupled  to the helical edges with amplitudes $w_{\sigma}$ \cite{rosenow} and also to an external reservoir of ordinary fermions with amplitude $w$. 
The Hamiltonian reads
\begin{eqnarray}\label{Htun}
H_{tun} &= &\sum_{\sigma} w_{\sigma} d^{\dagger}_{\sigma} \psi_{\sigma}(x=0) + w \sum_{\sigma,k} d^{\dagger}_{\sigma} c_{k \sigma}+\mbox{H.c.}.
\end{eqnarray}

\section{Linear thermoelectric description}

\subsection{Spin-dependent fluxes and Onsager matrix}
 We assume that the quantum dot is strongly coupled  to the external reservoir, so that it has the same chemical potential $\mu_2=\mu$ and temperature $T_2=T+ \Delta T$ of this system. The edge states are contacted
 to left and right reservoirs at chemical potential $\mu_1=\mu+ eV$ and $T_1=T$. 
 The charge and heat currents between the quantum dot and the helical edges is composed of  fluxes of $\uparrow$ electrons flowing to the $L$ channel and $\downarrow$ electrons
 flowing to the $R$ channel. We set $\mu=0$, and define these components as follows
 \begin{eqnarray}\label{cfluxes}
 J^C_{\sigma} &=&- 2 e w_{\sigma} \mbox{Re}\left[G_{d,\sigma}^<(t,t) \right],  \nonumber \\
 J^Q_{\sigma} & = &- 2 w_{\sigma} \mbox{Re} \left[i \partial_{t^{\prime}}G_{d,\sigma}^<(t,t^{\prime}) \right]_{t^{\prime}=t},
 \end{eqnarray}
 where we have introduced  the lesser Green's function $G_{d,\sigma}^<(t,t^{\prime}) = -i  \langle d^{\dagger}_{\sigma}(t) \psi_{\sigma}(t^{\prime}) \rangle$. In the next section we will explain the method to evaluate these
 currents. Here, we focus on small temperature differences $\Delta T$ and small bias voltages $eV$ between the reservoirs contacting the edges and the reservoir contacting the quantum dot.  Under this condition, these currents will
 be linear functions of the affinities $X_1= \Delta \mu/T$ and $X_2= \Delta T/k_B T^2$, with $\Delta \mu=\mu_2-\mu_1$ and $\Delta T=T_2-T_1$,
 \begin{eqnarray}\label{onsa}
 J^C_{\sigma} &=&L_{11}^{\sigma} X_1 + L_{12}^{\sigma} X_2, \nonumber \\
 J^Q_{\sigma} &=&L_{21}^{\sigma} X_1 + L_{22}^{\sigma} X_2.
 \end{eqnarray}
The coefficients $L_{ij}^{\sigma}$ obey Onsager relations $L_{ij}^{\uparrow}(B)=L_{ji}^{\downarrow}(-B)$. In addition, these coefficients satisfy constraints imposed by the second law of thermodynamics, according to which
the rate of entropy production is positive, \cite{casati, ora}
\begin{equation}\label{S-punto}
\dot{S}= \sum_{ i,j} X_i \; L_{ij} \; X_j \geq 0,
\end{equation}
with $L_{ij}=\sum_{\sigma}L_{ij}^{\sigma}$.
This implies $L_{11}, L_{22} \geq 0,\; \mbox{det}[L] \geq 0$.

The diagonal matrix elements define, respectively, the electrical and thermal conductances per spin channel, while the off-diagonal matrix elements define
the charge to energy conversion and vice versa. For $B \neq 0$, the charge current comes along with a spin current. In turn,  imbalance between the flows with $\uparrow$ and $\downarrow$ electrons implies a net spin current  flowing 
into the left or right terminal of the QSH bar.   Therefore, we can identify different thermoelectric operational modes for this setup, which we describe below. 

\subsection{Operational modes}\label{oper}
The different interesting operational modes are: charge heat engine or refrigerator, spin heat engine and spin refrigerator. They are  illustrated in Figs. \ref{fig1a} and \ref{fig1}. While the charge heat engine or refrigerator mode
can be realized with magnetic field as well as without magnetic field, the spin heat engine and spin refrigerator modes operate only with a finite magnetic field. The conditions of operation corresponding to  each case are indicated in section V. 

\subsubsection{Charge heat engine or refrigerator}\label{oper-1}

\begin{figure}[h]
\vspace{1.cm}
\begin{center}
\includegraphics[width=\columnwidth]{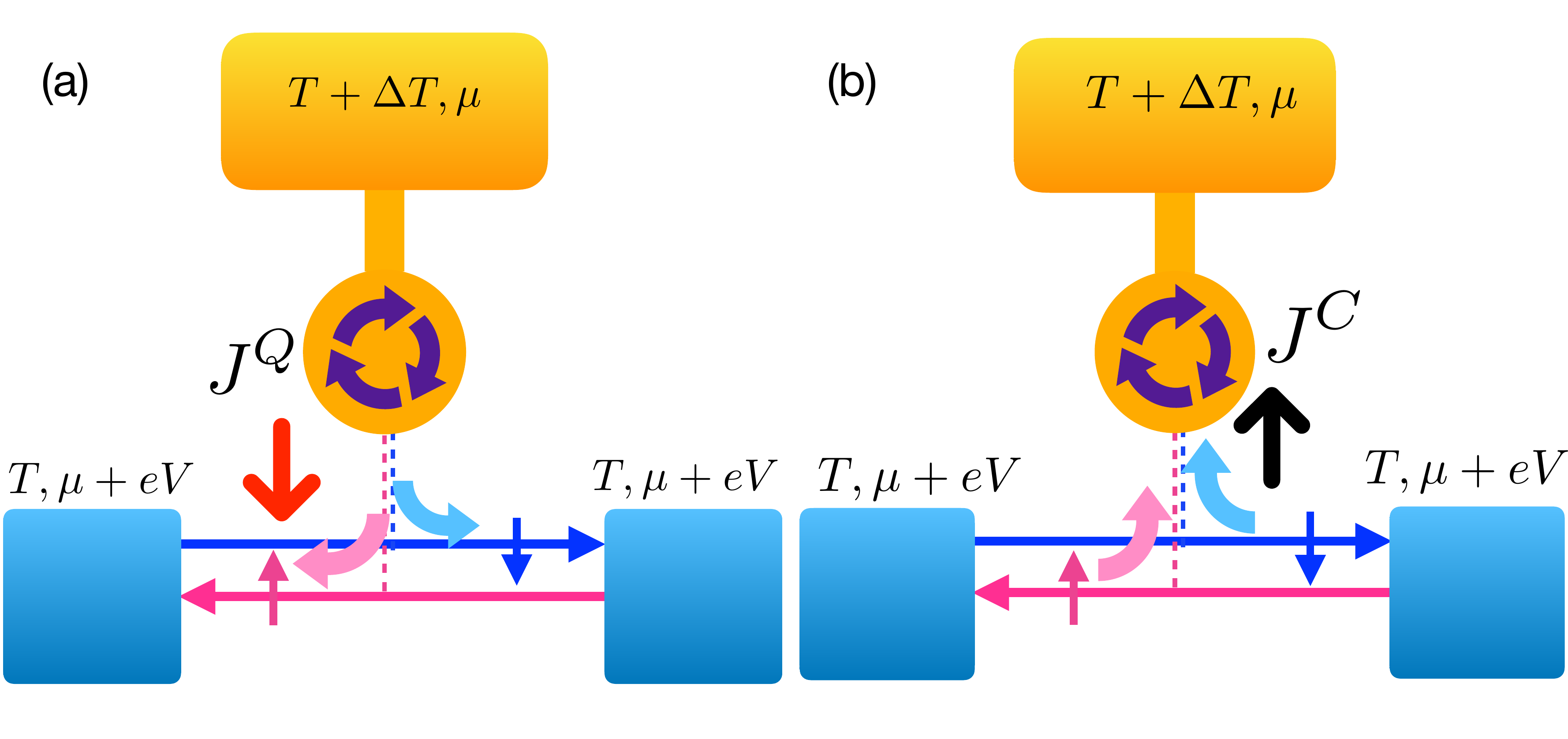}
\end{center}
\caption{(a) Charge heat engine. The heat current from the hottest reservoir is converted into charge current flowing through the tunneling contact and equally distributed, in absence of magnetic field, to the left and right reservoir of the QSH system.
(b) Charge refrigerator. An electric current flows from the edge states to the quantum dot and is accompanied by a heat current that refrigerates the left and right reservoirs of the QSH system. }
\label{fig1a}
\end{figure}
We start by considering the cases sketched  in Fig. \ref{fig1a}. These are the usual operational modes of a two-terminal configuration with chemical potential and a temperature differences  in opposition.  In our case, the hot terminal is the quantum dot connected to the reservoir at $\mu$  and
$T+ \Delta T$, while the cold reservoir is the Luttinger liquid composed of the two helical edge states. The thermoelectric response can be described by
\begin{equation}\label{charge}
\left(\begin{array}{c}
J^C \\
J^Q 
\end{array}
\right)= \left(\begin{array}{cc}
{\cal L}^c_{11} & {\cal L}^c_{12}\\
{\cal L}^c_{21} & {\cal L}^c_{22}\end{array} \right) \left(\begin{array}{c}
X_1 \\
X_2 
\end{array}
\right).
\end{equation}
In terms of the fluxes and Onsager coefficients defined in Eq. (\ref{onsa}) we have $J^C= \sum_{\sigma} J^C_{\sigma} $, $J^Q= \sum_{\sigma} J^Q_{\sigma} $ and ${\cal L}^c_{ij} = \sum_{\sigma} L^{\sigma}_{ij}$. The device can operate as a heat
engine or refrigerator.  
The 
performance of these operational modes
is qualified by the efficiency (heat engine) or coefficient of performance (refrigerator) as follows
\begin{equation}
\eta^{c,he}= - \frac{J^C X_1}{J^Q}, \;\;\;\;\;\;\;\;\;\;\;\;\;\;\eta^{c,fri}= -\frac{J^Q}{J^C X_1}.
\end{equation}
In the former case $J^C$ flows against the bias voltage $X_1$ and a electrical power $-J^C X_1$ is generated at expenses of  the investment of a heat flow $J^Q$, while in the latter case
a heat current $-J^Q$  is extracted from the coldest reservoir by investing an electrical power $J^C X_1$. Both coefficients are bounded by the Carnot limits, $\eta^{c,he} \leq \Delta T/T$ and 
$\eta^{c,fri} \leq T/\Delta T$.

\subsubsection{Spin heat engine}\label{oper-2}

\begin{figure}[h]
\vspace{1.cm}
\begin{center}
\includegraphics[width=\columnwidth]{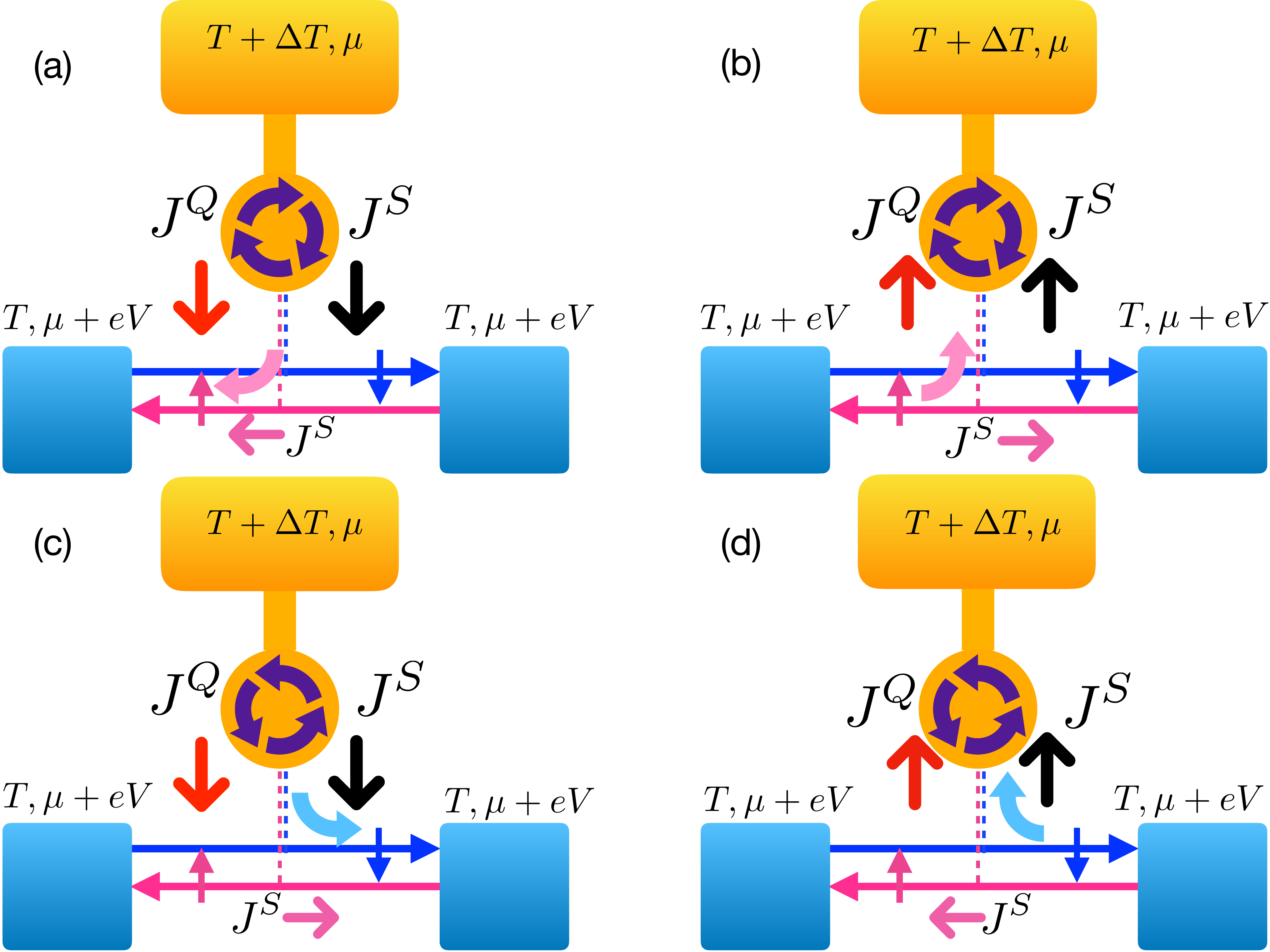}
\end{center}
\caption{(a) and (c) Spin heat engine. The heat current from the hottest reservoir is converted into a spin current  flowing through the tunneling contact and to the left (for $\uparrow$ electrons) and right (for $\downarrow$ electrons) reservoir of the QSH system.
(b) and (d) Spin refrigerator. A polarized ($\uparrow$ or $\downarrow$) electric  current flows from the edge states to the quantum dot  and is accompanied by a refrigeration of  the left or right reservoirs of the QSH system.}
\label{fig1}
\end{figure}

We now consider the cases sketched  in Fig. \ref{fig1} (a) and (c). The magnetic field at the quantum dot filters electrons with a given spin component. The heat current flowing from the hot reservoir in strong coupling to the quantum dot leads to 
a polarized electron current through the tunneling contact to the helical edge states and a spin current is induced at the QSH system. Due to the helical nature of the edge states, this also implies a charge current flowing to the left or to the right along the edge.
Hence the direction of the magnetic field distributes the electron flow towards the left or right terminals of the QSH bar. 
The thermoelectric description to characterize the generation of a spin current  $J^S$ in the QSH by using the heat current  $J^Q$ 
can be formulated in terms of the linear dependence of these currents with the affinities,
 
\begin{equation}\label{spin}
\left(\begin{array}{c}
J^S \\
J^Q 
\end{array}
\right)= \left(\begin{array}{cc}
{\cal L}^s_{11} & {\cal L}^s_{12}\\
{\cal L}^c_{21} & {\cal L}^c_{22}\end{array} \right) \left(\begin{array}{c}
X_1 \\
X_2 
\end{array}
\right),
\end{equation}
with $J^S= \xi \left( J^C_{\uparrow} -J^C_{\downarrow} \right)$ where $\xi=\mbox{sgn}\left( L^{\uparrow}_{11} - L^{\downarrow}_{11} \right)$. 
The matrix elements of Eq. (\ref{spin}) in terms of the original Onsager coefficients read ${\cal L}^s_{1j} = \xi \left( L^{\uparrow}_{1j} - L^{\downarrow}_{1j} \right)$ and ${\cal L}^c_{2 j} $. We have introduced the sign $\xi$ in the
definition, in order to have ${\cal L}^s_{11} >0$ and $J^Q$ is the total heat current previously defined.
In this case, the efficiency is quantified as the ratio between fraction of the electrical power flowing against the bias voltage to the left or right terminals (for $\xi >0$ or $\xi <0$), $-J^S X_1$,
and the heat current flowing from the hot reservoir $J^Q$ 
\begin{equation}\label{eficiencia-maquina-spin}
\eta^{s,he}= - \frac{J^S X_1}{J^Q}.
\end{equation}
Notice that $J^S X_1$ has the same units as the electrical power, since $J^S$ is a polarized charge current.

\subsubsection{Spin refrigerator}\label{oper-3}

Finally we consider the cases  sketched Fig. \ref{fig1} (b) and (d). They can be regarded as the reversed operational mode of the spin heat engine. As a consequence of the spin filtering introduced by the magnetic field, not only a
polarized electron flux  but also an associated energy flux is established between the left and right terminals of the QSH bar. This can be  described by the heat current $J^{Q,s}= \xi \left( J^Q_{\uparrow} -J^Q_{\downarrow} \right)$, with $\xi=\mbox{sgn}\left( L^{\uparrow}_{22} - L^{\downarrow}_{22}\right)$. Hence, a polarized charge flux from the helical edges towards the quantum dot
and the external terminal can be used to refrigerate the left or the right reservoir contacting the QSH system. 
The thermoelectric response can be described as follows
\begin{equation}\label{spinq}
\left(\begin{array}{c}
J^C \\
J^{Q,s} 
\end{array}
\right)= \left(\begin{array}{cc}
{\cal L}^c_{11} & {\cal L}^c_{12}\\
{\cal L}^s_{21} & {\cal L}^s_{22}\end{array} \right) \left(\begin{array}{c}
X_1 \\
X_2 
\end{array}
\right),
\end{equation}
The matrix elements of Eq. (\ref{spinq}) in terms of the original Onsager coefficients read 
${\cal L}^s_{2j} = \xi \left( L^{\uparrow}_{2j} - L^{\downarrow}_{2j} \right)$ and the efficiency can be quantified as the heat current extracted from the left or the right reservoirs (for $\xi >0$ or $\xi <0$), upon
investing a total electrical power $J^C X_1$,
\begin{equation}\label{eta-s-fr}
\eta^{s,fr}= - \frac{J^{Q,s}}{J^C X_1}.
\end{equation}

\section{Calculation of the Onsager coefficients}

The spin-dependent charge and heat currents defined in Eqs. (\ref{cfluxes}) can be calculated by means of non-equilibrium Schwinger Keldysh formalism and treating perturbatively the tunneling couplings between the  quantum dot and the helical edges, 
$w_{\sigma}$ . Here we summarize the main steps of this calculation. 
The charge current can be expressed as follows
\begin{eqnarray}\label{I-idot-2}
 J_{\sigma}^C & = & -2e w_{\sigma}^2  \int^{+\infty}_{-\infty}dt \sum_{s=\pm} \mbox{Re} \left[ g^{<}_{d,s }(t)u^{2}_{s,\sigma} g^{>}_{\sigma}(-t) \right. \nonumber \\
                                                                  & & \left.\;\;\;\;\;\; - g^{>}_{d, s }(t) u^{2}_{s,\sigma} g^{<}_{\sigma}(-t) \right],
\end{eqnarray}
where $u_{s,\sigma}$ are defined in Eq. (\ref{change}).
The lesser and greater Green's functions of the quantum dot contacted only to the ordinary reservoir read $g^{>, <}_{d s} (t) = \int d \omega e^{-i \omega/\hbar t} \lambda^{>, <}(\omega) \rho_{d s }(\omega)$.
They depend on the density of states of the quantum dot coupled to the external reservoir,
\begin{equation}
\rho_{d s }(\omega)=   \frac{ 1}{\pi}\frac{\gamma}{\omega - E_{s} + i \gamma}.
\end{equation}
 The functions $\lambda^{<}(\omega) = i f_d(\omega)$ and $\lambda^{>}(\omega) = -i \left[ 1- f_d(\omega) \right]$, depend on the chemical potential $\mu$ and temperature $T+ \Delta T$ of the 
 external reservoir in strong coupling with the quantum dot through the Fermi function
 $f_d(\omega)$. The functions $g^{>,<}_{\sigma}(t)$ correspond to the left moving (for $\sigma=\uparrow$) and right moving (for $\sigma=\downarrow$) electrons at the edge states. They can be written as
$g^{>}_{\sigma}(t)=-g^{<}_{\sigma}(-t)$, with
\begin{eqnarray}\label{rho-lutt}
g^{<}_{\sigma} (t) & = & \frac{i}{h} \int d \omega e^{-i \omega/\hbar t} \rho_{\sigma}(\omega) f_{\sigma}(\omega+ eV),  \\
\rho_{\sigma}(\omega)&=&a^{\bar{K}-1} \frac{(2\pi T)^{\bar{K}-1}}{\Gamma(\bar{K})} 
              \Big\vert \frac{\Gamma(\bar{K}/2 + i\omega/2\pi T)}{\Gamma(1/2 + i\omega/2\pi T)}\Big\vert^2.
\end{eqnarray}
 $ \rho_{\sigma}(\omega)$ is the density of states of the chiral Luttinger liquid of left (for $\uparrow$) and right (for $\downarrow$) movers, 
 $\Gamma(x)$ is the Gamma function and 
 \begin{equation}\label{overk}
 \overline{K}=(K+1/K)/2.
 \end{equation}
  More details on this calculation are presented in Appendix \ref{green}.
 
 Substituting these Eqs. in Eq. (\ref{I-idot-2}) we get the following expression for the charge current through the tunneling contact 
 \begin{equation}\label{I-idot-5}
J_{\sigma}^C =  \frac{e}{h} \int^{+\infty}_{-\infty}d\omega {\cal T}_{\sigma}(\omega)
 \Big[f_{\sigma}(\omega + eV )-f_d(\omega) \Big] ,\nonumber\\
\end{equation}
with 
\begin{equation}
{\cal T}_{\sigma}(\omega)= 4\pi  w^{2}_{\sigma} \sum_s u^{2}_{s,\sigma} \rho_{d, s}(\omega)\rho_{\sigma}(\omega + e V).
\end{equation}
Similarly, for the heat current we get
 \begin{equation}\label{I-idot-6}
J_{\sigma}^Q =  \int^{+\infty}_{-\infty}d \omega ~\frac{\omega}{h}  {\cal T}_{\sigma}(\omega)\Big[f_{\sigma}(\omega + e V)-f_d(\omega) \Big].
\end{equation}
Expanding these expressions up to linear order in $eV$ and $\Delta T$ we get the Onsager coefficients of Eq. (\ref{onsa}), 

\begin{equation} \label{coeff}
\widehat{L}^{\sigma}=-\frac{k_B T}{2 h} \int d \omega \left(\begin{array}{cc} 
e & e \omega \\
\omega & \omega^2 \end{array} \right) {\cal T}_{\sigma}(\omega) \; \frac{\partial f(\omega)}{\partial \omega}.
\end{equation}

In the limit of low temperature $k_B T < \gamma$, the density of states of the
Luttinger liquid can be well approximated by a power law as in Ref. \onlinecite{enhan}. The resulting expressions for the Onsager coefficients  in this limit are
\begin{eqnarray}\label{coeff-approx}
L_{11}^{\sigma}& \simeq & T \sum_s u^{2}_{s,\sigma}\rho_{d, s }(0) I_0 (\overline{K}), \nonumber \\
L_{12}^{\sigma}& \simeq & T\sum_s u^{2}_{s,\sigma} \rho^{\prime}_{d,s}(0) I_2 (\overline{K}),        \;\;\;\;L_{21}^{\sigma} \simeq L_{12}^{\sigma}, \nonumber \\
L_{22}^{\sigma} & \simeq & T\sum_s u^{2}_{s,\sigma} \rho_{d, s}(0) I_2 (\overline{K}),
\end{eqnarray}
with
\begin{equation}\label{integrales}
I_n(\overline{K})= c(\overline{K},T) \frac{(k_B T)^{\overline{K}-1+n}}{\overline{K}-1+n},
\end{equation}
where $c(\overline{K},T)$ is a coefficient depending on $\overline{K}$ and $T$.  Interestingly, from the definition of Eq. (\ref{overk}) we see that  $\overline{K}$  is a function of the Luttinger parameter $K$ obeying the following
symmetry $K \leftrightarrow 1/K$. Hence, $\overline{K}$ has the same behavior for repulsive ($K<1$) and attractive ($K>1$) interactions. This is because the infinite Luttinger liquid for the pair of edge states defined in Eq. (\ref{Hll})
can be described in terms of 
two  Hamiltonians, which are bilinear in bosonic fields resulting from combinations of the original ones $\phi_{\sigma}$. These two Hamiltonians are related by a
duality transformation  under the change $K \leftrightarrow 1/K$, being the non-interacting case, $K=1$ self-dual. 
The local tunneling density of states depends on the two combinations of bosonic fields, hence, on the parameter $\overline{K}$ defined in
Eq. (\ref{overk}). \cite{kane-fish-92}

\section{Results}

\subsection{Charge heat engine and refrigerator}

In the operational modes introduced in Section \ref{oper-1} it is possible to proceed as in the usual $2 \times 2$ thermoelectric devices  \cite{casati} and parametrize the efficiency by a figure of merit 
\begin{equation}
ZT= \frac{({\cal L}^c_{12})^2}{\mbox{Det}{\widehat{\cal L}}}, 
\end{equation}
being ${\cal L}^c_{12}={\cal L}^c_{21}$. $\widehat{\cal L}$ is the Onsager matrix characterizing the corresponding operational mode and ${\cal L}^c_{ij}$ the corresponding matrix elements. Following Ref. \onlinecite{casati}, we
 can choose $\Delta \mu=-eV$ and $\Delta T>0$, and derive  the maximum efficiency 
for a fixed $\Delta T$. It
can be expressed as
\begin{equation}
\eta^{\rm max} = \eta_C \frac{ \sqrt{ZT + 1} -1}{\sqrt{ZT + 1} +1},
\end{equation}
being $\eta_C= \Delta T/T$ the Carnot efficiency for the heat engine regime and $\eta_C = T/\Delta T$ the Carnot coefficient of performance for the refrigerator regimes. 
This maximum corresponds to the following relation between $\Delta \mu$ and $\Delta T$,
\begin{equation}
T \Delta \mu= - \Delta T \frac{{\cal L}_{22}}{{\cal L}_{12}} \left(1- \sqrt{\frac{\mbox{Det}{\widehat{\cal L}}}{{\cal L}^c_{11} {\cal L}^c_{22}}} \right).
\end{equation}
The device operates as a heat engine within the range of voltages satisfying
\begin{eqnarray}\label{he}
& & -\frac{{\cal L}^c_{12}}{ {\cal L}^c_{11}} \Delta T \leq T \Delta \mu \leq 0,\;\;\;\;\;\;\;\;\;\;\;\;\;{\cal L}^c_{12} >0,\nonumber \\
& & 0 \leq T \Delta \mu\leq  -\frac{{\cal L}^c_{12}}{ {\cal L}^c_{11}}  \Delta T,\;\;\;\;\;\;\;\;\;\;\;\;\;{\cal L}^c_{12} <0,
\end{eqnarray}
while it operates as a refrigerator within the range 
\begin{eqnarray}\label{fr}
& &  T \Delta \mu <-\frac{{\cal L}^c_{22}}{ {\cal L}^c_{21}}  \Delta T,\;\;\;\;\;\;\;\;\;\;\;\;\;{\cal L}^c_{21} >0,\nonumber \\
& &  T \Delta \mu >  -\frac{{\cal L}^c_{22}}{ {\cal L}^c_{21}}  \Delta T,\;\;\;\;\;\;\;\;\;\;\;\;\;{\cal L}^c_{21} <0.
\end{eqnarray}
In what follows, we show and analyze the Onsager coefficients and the figure of merit described in the previous section. We calculate numerically the coefficients of Eq. (\ref{coeff}) and derive some analytical results based on
Eqs. (\ref{coeff-approx}) within the
low-temperature regime $k_B T < \gamma$.
 We have verified that the latter are in prefect agreement with the exact results.

\subsubsection{$B=0$}
\begin{figure}[tbp]
\vspace{1.cm}
\begin{center}
\includegraphics[width=\columnwidth]{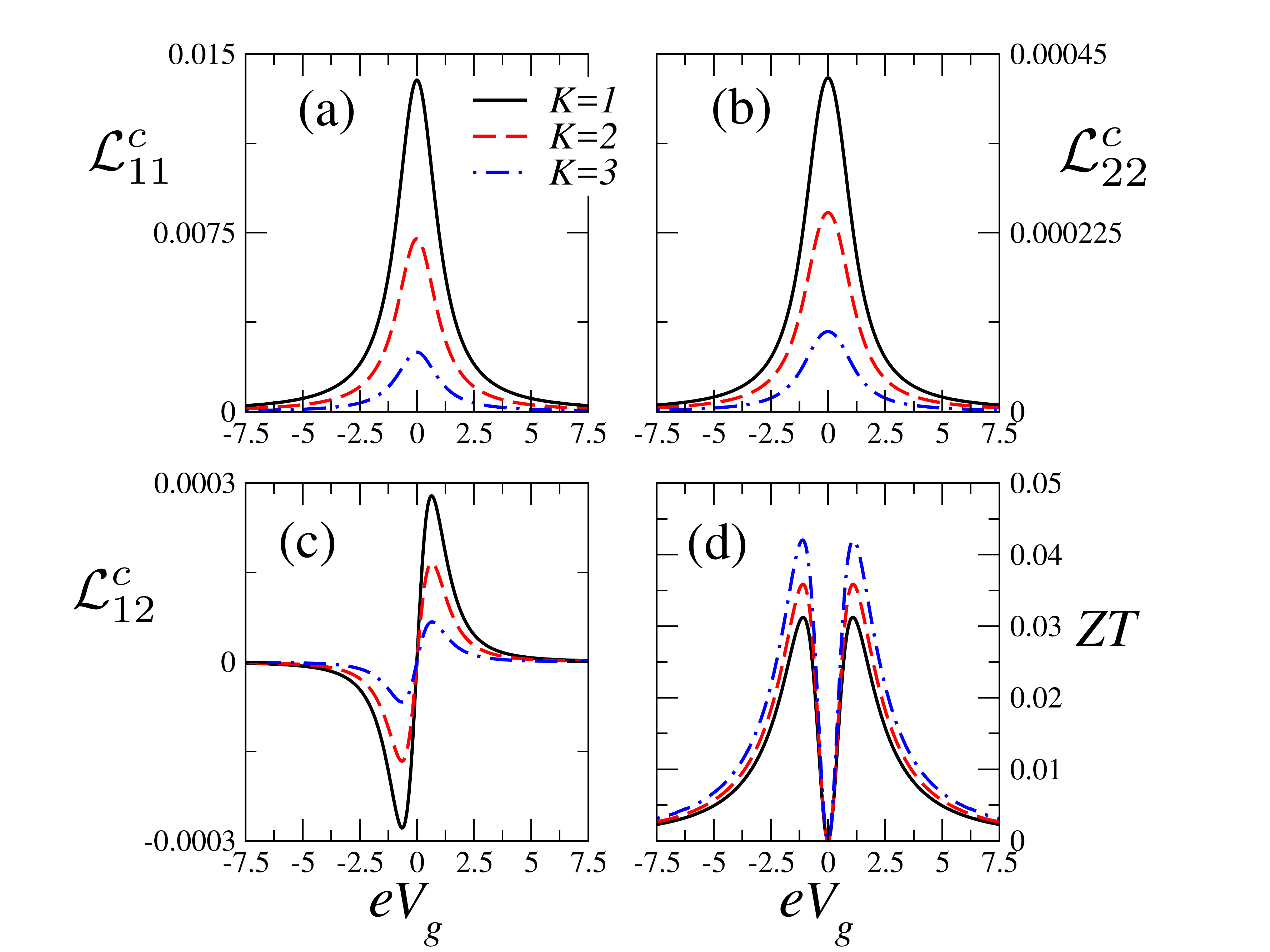}
\end{center}
\caption{(Color online) Onsager coefficients (panels (a), (b) and (c)) and figure of merit $ZT$ (panel (d)) 
corresponding at the charge heat engine or refrigerator
at the temperature $k_B T=0.1\gamma$ and $B=0$, as functions of the gate 
voltage (in units of $\gamma/e$). The unit of ${\cal L}^c_{11}, \;{\cal L}^c_{12},\;{\cal L}^c_{22}$ are, respectively,
$ e \gamma     / h,\;
e \gamma^2 / h,\;\gamma^3 / h$.
Solid, dashed and dashed-dotted lines correspond to $K=1, 2, 3$, respectively. }
\label{regimen-1a}
\end{figure}

In Fig. \ref{regimen-1a}, we show the Onsager coefficients for the charge heat engine or refrigerator described in Section \ref{oper-1}, as well as the corresponding figure of merit. We fix the magnetic field at $B=0$ 
a low temperature $T=0.1\gamma$, and analyze these quantities  as functions of the gate 
voltage.

 The diagonal coefficients ${\cal L}^{c}_{11}$ and ${\cal L}^{c}_{22}$, related, respectively, to the electrical and thermal conductances, have maxima when the level of the dot
is aligned with the chemical potential $\mu$.
The maximum thermoelectric efficiency, corresponding to the maximum $ZT$, takes place at the maximum of the absolute value of the off-diagonal coefficient $|{\cal L}^c_{12}|$. This is achieved by applying gate voltages leading to configurations with a high density of states of the quantum dot but with broken particle-hole symmetry. This condition is satisfied within a window $ e|V_g| \sim  \gamma $. 
The sign of ${\cal L}^{c}_{12}={\cal L}^{c}_{21}$ defines the range of voltages for the operational mode, as discussed in Eqs. (\ref{he}) and (\ref{fr}).

Comparing to the  fractional quantum Hall regime, modeled by a chiral Luttinger liquid, we notice that  the inverse of the fractional filling factor $1/\nu$ plays a similar role 
as the parameter $\overline{K}$ defined in Eq. (\ref{overk})
in the present case. Therefore, akin to the fractional quantum Hall effect analyzed in Ref. \onlinecite{enhan},
 the thermoelectric performance is enhanced when the system departs from the non-interacting limit $K=1$.  This is true for $K>1$ as well as $K<1$. 

\subsubsection{Effect of the magnetic field}
\begin{figure}[hbt]%[tbp]
\vspace{1.cm}
\begin{center}
\includegraphics[width=\columnwidth]{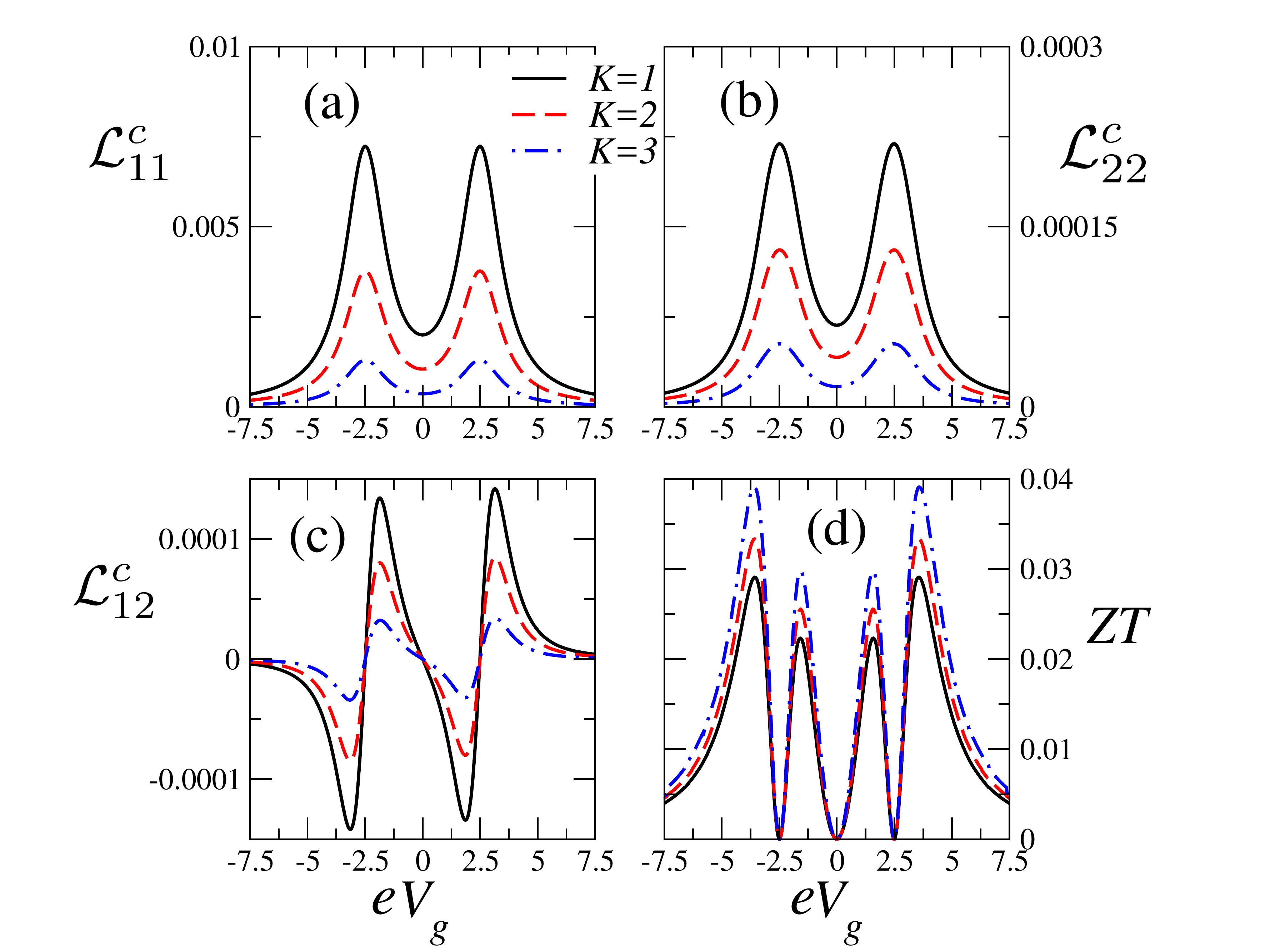}
\end{center}
\caption{Onsager coefficients (panels (a), (b) and (c)) and figure of merit $ZT$ (panel (d)) 
corresponding at the charge heat engine or refrigerator
at the temperature $k_B T=0.1\gamma$ and magnetic field $B=5\gamma/\mu_B$ and $\theta=0$, as functions of the gate 
voltage.  Other details are the same as in the previous Fig. }
\label{regimen-1b}
\end{figure}

Although the magnetic field is not essential for the device to operate as a charge heat engine or refrigerator, it is interesting to analyze its effect.  
Turning on the magnetic field, $B$, the electronic levels of the quantum dot is split by Zeeman effect into two levels  with energies $E_{\pm}$ given in Eq. (\ref{eigenvalue}). 
Figure (\ref{regimen-1b}) shows the thermoelectric coefficients and the figure of merit for a finite magnetic field aligned with the direction of the spin-orbit coupling of the QSH bar, $\theta=0$. We can identify in the  behavior of the diagonal coefficients 
${\cal L}^{c}_{11}$ and ${\cal L}^{c}_{22}$  as functions of the gate voltage (top panels) the two peaks at $eV_g = \pm \mu_B B/2$. These correspond to the values of the gate voltage for which  the levels are aligned with the mean
chemical potential $\mu$. Concomitantly, the non diagonal coefficient ${\cal L}^{c}_{12}$ also presents more features than in the case of vanishing magnetic field. In particular, this coefficient vanishes at the two resonant values 
$eV_g = \pm \mu_B B/2$, in addition to the particle-hole symmetric value $e V_g=0$ of the zero magnetic field.

Changing the direction of the applied magnetic field, $\theta$, has no effect on the Onsager coefficients ${\cal L}^{c}_{ij}$.  This result  can be understood on the basis of the expressions given in Eq. (\ref{coeff-approx}). 
Considering, for instance, the case of the low-temperature expression for the coefficient
\begin{eqnarray}\label{theta-independent}
{\cal L}^{c}_{11}&=&\sum_{\sigma}L_{11}^{\sigma}\simeq T \sum_{s,\sigma} u^{2}_{s,\sigma}\rho_{d, s }(0) I_0 (\overline{K})\nonumber\\
                 &=&T \sum_{s} \rho_{d, s }(0) I_0 (\overline{K}),
\end{eqnarray}
and taking into account that from Eq. (\ref{change})  $\sum_{\sigma}u^{2}_{s,\sigma}=1$, we conclude that the component $B_{\perp}$ of the magnetic field does not play any role in this regime. This is also true for the exact coefficients evaluated numerically.
Hence, the perfect alignment of the magnetic field along the direction of the spin-orbit  axis of the topological insulator, is not crucial for the thermoelectric performance of this operational mode. 

\begin{figure}[tbp]
\vspace{1.cm}
\begin{center}
\includegraphics[width=0.8\columnwidth]{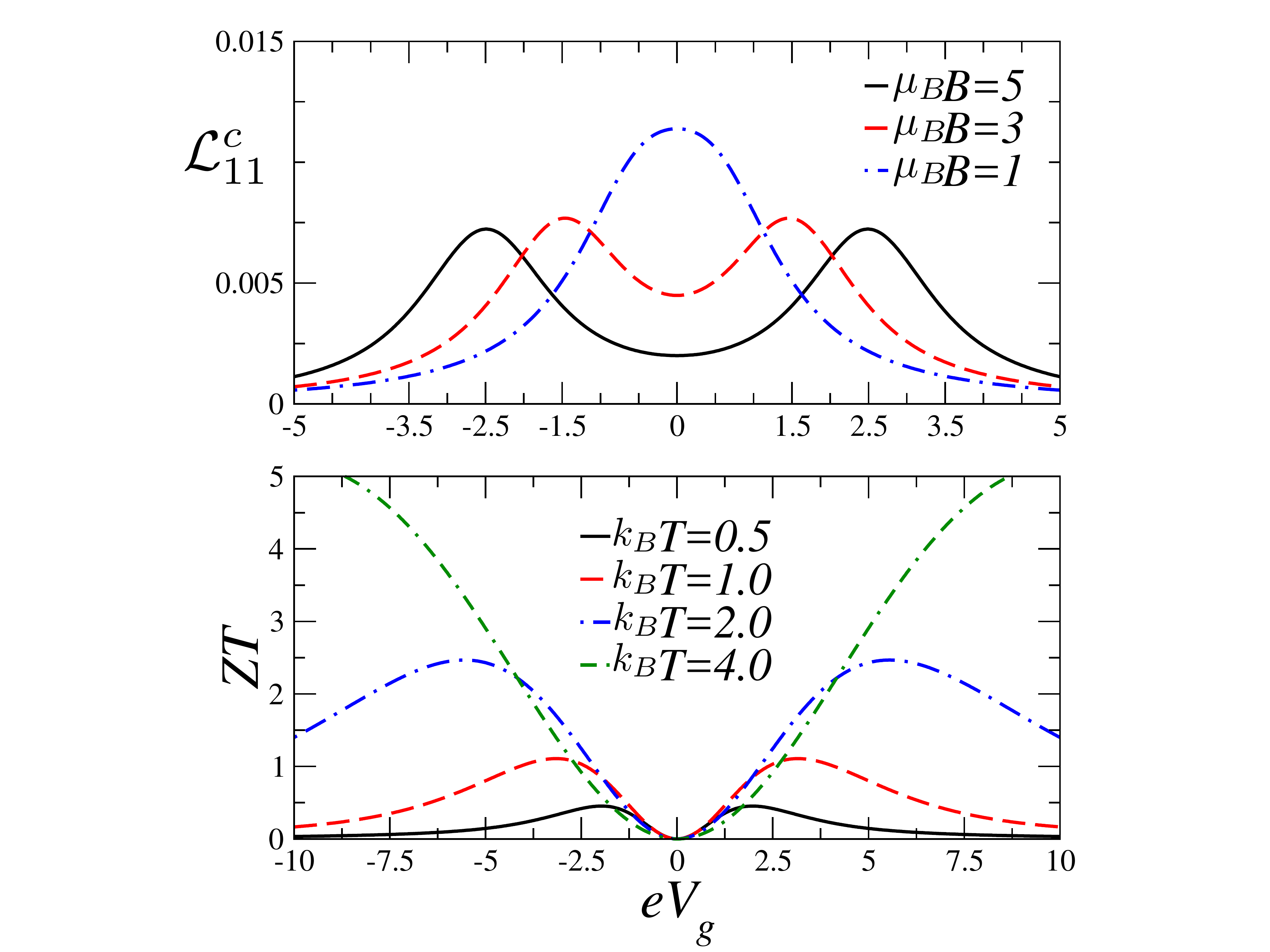}
\end{center}
\caption{(Color online) Top panel: Diagonal coefficient ${\cal L}^{c}_{11}$ at the temperature $k_B T=0.1\gamma$ and $\theta=0$ and $K=1$ for several values of the magnetic field as a function of the gate voltage. Lower panel: Figure of merit at zero magnetic field for several values of temperature and $K=1$.  The operational mode is charge heat engine or refrigerator. Other details are the same as in the previous Fig.}
\label{regimen-1c}
\end{figure}

In the top panel of Fig. (\ref{regimen-1c}) we show the behavior of ${\cal L}^{c}_{11}$ for several values of the magnetic field.
When the magnitude of the magnetic field is reduced  we observe a decreasing resolution of the peaks located at the energies of the dot in the diagonal coefficients.
Both peaks merge into a single peak for values of the applied field approaching to the intrinsic broadening $\gamma$  of the quantum dot levels. The effect of the temperature is analyzed in the lower panel of Fig. (\ref{regimen-1c}). As 
the temperature is increased, large values of the figure of merit are achieved. This is 
similar to the results for the fractional quantum Hall effect presented in Ref. (\onlinecite{enhan}). Substituting the low-temperature behavior for the Onsager coefficients given by Eqs. (\ref{coeff-approx}) we get the following analytical behavior for the low-temperature regime
\begin{eqnarray}\label{ztlt}
ZT &\sim& \left(\frac{\rho^{\prime}_d(0)}{\rho_d(0)}\right)^2 \frac{\overline{K}^2-1}{(\overline{K}+1)^2} (k_B T)^2,\;\;\;\;K \neq 1\nonumber, \\
ZT &\sim& \left(\frac{\rho^{\prime}_d(0)}{\rho_d(0)}\right)^2  (k_B T)^2,\;\;\;\;\;\;\;\;\;\;\;\;\;\;\;\;\;\;\; K = 1.
\end{eqnarray}
The above expression is in perfect agreement with the analytical results.

\subsection{Spin heat engine}
For the spin heat engine mode described in section \ref{oper-2}, we are interested in the conversion of the heat current from the external reservoir to a spin current through the quantum dot and along the edge states of the QSH bar.
This mode can only be implemented by applying a magnetic field at the quantum dot. 
To characterize it, we introduced  the coefficients ${\cal L}^{s}_{11}$ and ${\cal L}^{s}_{12}$ in Eq. (\ref{spin}) to describe the linear dependence of the induced spin current as a function of the affinities $X_1,\; X_2$. 

\begin{figure}[tbp]
\vspace{1.cm}
\begin{center}
\includegraphics[width=\columnwidth]{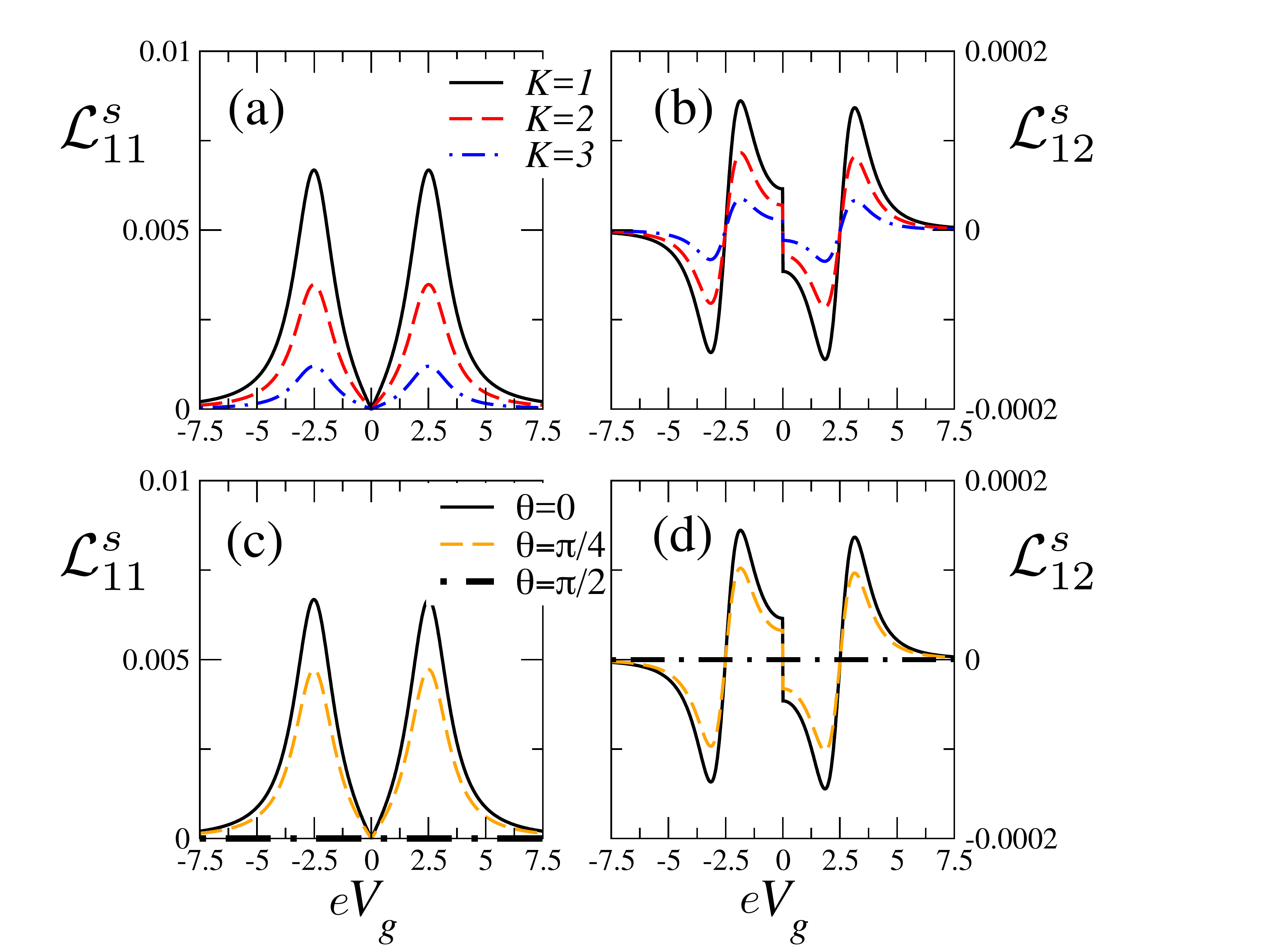}
\end{center}
\caption{(Color online) Onsager coefficients ${\cal L}^s_{11}$ and ${\cal L}^s_{12}$  
corresponding at the spin heat engine
at the temperature $k_B T=0.1\gamma$ and magnetic field $\mu_B B=5\gamma$, as functions of the gate 
voltage. Top panels show the $K$ dependence for $\theta=0$. Lower panels show the $\theta$ 
dependence for $K=1$. Other details are the same as in the previous Fig. }
\label{regimen-2a}
\end{figure}

The efficiency of the device can be quantified by the ratio between the power developed by the polarized current $J^S$ and the heat flux $J^Q$, as defined in Eq. (\ref{eficiencia-maquina-spin}).  Following the same philosophy 
as  in the usual charge heat engine and refrigerator, we  focus on fixed $\Delta T$ and analyze the conditions for the maximum possible efficiency within the linear response regime. To this end, we maximize 
Eq. (\ref{eficiencia-maquina-spin}) for fixed $\Delta T$. The maximum value is 
\begin{equation} \label{maxshe}
\eta^{s,he,{\rm max}}= \eta_c \frac{ \chi^{he} \sqrt{1 + (ZT)^{s,he}} - 1}{\sqrt{1 + (ZT)^{s,he}} + 1}.
\end{equation}
We have used the following definitions
\begin{equation}\label{chi}
\chi^{he}= \frac{{\cal L}_{12}^s}{ {\cal L}_{21}^c}, \;\;\;\;\;\;\;\;\;\;\;\; (ZT)^{s,he}  = \frac{{\cal L}_{12}^s {\cal L}_{21}^c }{ \mbox{Det}\widehat{\cal L}^{s,he}},
\end{equation}
being $\mbox{Det}\widehat{\cal L}^{s,he}$  the determinant of the matrix of Eq. (\ref{spin}). The maximum is
found to take place at the value of the voltage bias
\begin{equation}\label{tdmu}
 T \Delta \mu = - \Delta T \; \frac{{\cal L}_{12}^s}{{\cal L}_{11}^s R^{he}} \left( 1-\sqrt{1-R^{he}} \right),
\end{equation} 
being $R^{he}=\frac{{\cal L}_{12}^s {\cal L}_{21}^c}{ {\cal L}_{11}^s{\cal L}_{22}^c}$.
The maximum efficiency at fixed $\Delta T$ is parametrized by a figure of merit $(ZT)^{s,he}$, which has the same formal expression as for the charge heat engine. The maximum is achieved for parameters satisfying
$(ZT)^{s,he} \rightarrow \infty $ and it is bounded by $\chi^{he} \, \eta_c$. 
Notice that  $\chi^{he} \leq 1$,  and the limit $\chi^{he} =1$ corresponds to a fully polarized current. In addition to the definitions of Eqs. (\ref{maxshe}), (\ref{chi}) and
(\ref{tdmu}), it is important to take into account that for the spin heat engine operational mode to take place, it is necessary to satisfy the following conditions 
\begin{eqnarray}\label{she}
& & -\frac{{\cal L}^s_{12}}{ {\cal L}^c_{11}} \Delta T \leq T \Delta \mu \leq 0,\;\;\;\;\;\;\;\;\;\;\;\;\;{\cal L}^s_{12} >0,\nonumber \\
& & 0 \leq T \Delta \mu\leq  -\frac{{\cal L}^s_{12}}{ {\cal L}^c_{11}} \Delta T,\;\;\;\;\;\;\;\;\;\;\;\;\;{\cal L}^s_{12} <0,
\end{eqnarray}
From figures (\ref{regimen-1b}) and (\ref{regimen-2a}) we can observe that ${\cal L}_{21}^c$ has the same sign of ${\cal L}_{12}^s$ and therefore the ratio $R^{he}$ is a positive magnitude which is bounded by $0<R^{he}<1$ due to the constraints imposed by $\mbox{Det}\widehat{\cal L}^{s,he}>0$. This determines the selected sign for the root in Eq. (\ref{tdmu}).
In addition, we recall that the coefficients $L_{ij}^{\sigma}$ should satisfy the conditions leading to a positive rate of the entropy production given by Eq. (\ref{S-punto}).

Fig. (\ref{regimen-2a}) shows  the transport coefficients for a finite magnetic field aligned along the direction of the spin-orbit direction in the bar, $\theta=0$. 
The overall behavior of these coefficients and their dependence with the interactions $K$, are similar to  the charge heat engine (refrigerator) mode. From the definition of the sign $\xi$ in Eq. (\ref{spin}), the coefficient 
${\cal L}^{s}_{11}$ has non negative values and the peaks are at the energies $eV_g = \pm B/2$.  The different signs of $\xi$ imply different polarizations of the spin current, hence, different directions of the current along the 
QSH edge.
The coefficient  ${\cal L}^{s}_{12}$  does not continuously cross zero as the gate voltage pass from negative to positive values. This is explicitly seen in the discontinuity of ${\cal L}^{s}_{12}$ at $eV_g=0$,
as a consequence  of the sign $\xi$ imposed by ${\cal L}^{s}_{11}$.

\begin{figure}[tbp]
\vspace{1.cm}
\begin{center}
\includegraphics[width=0.8\columnwidth]{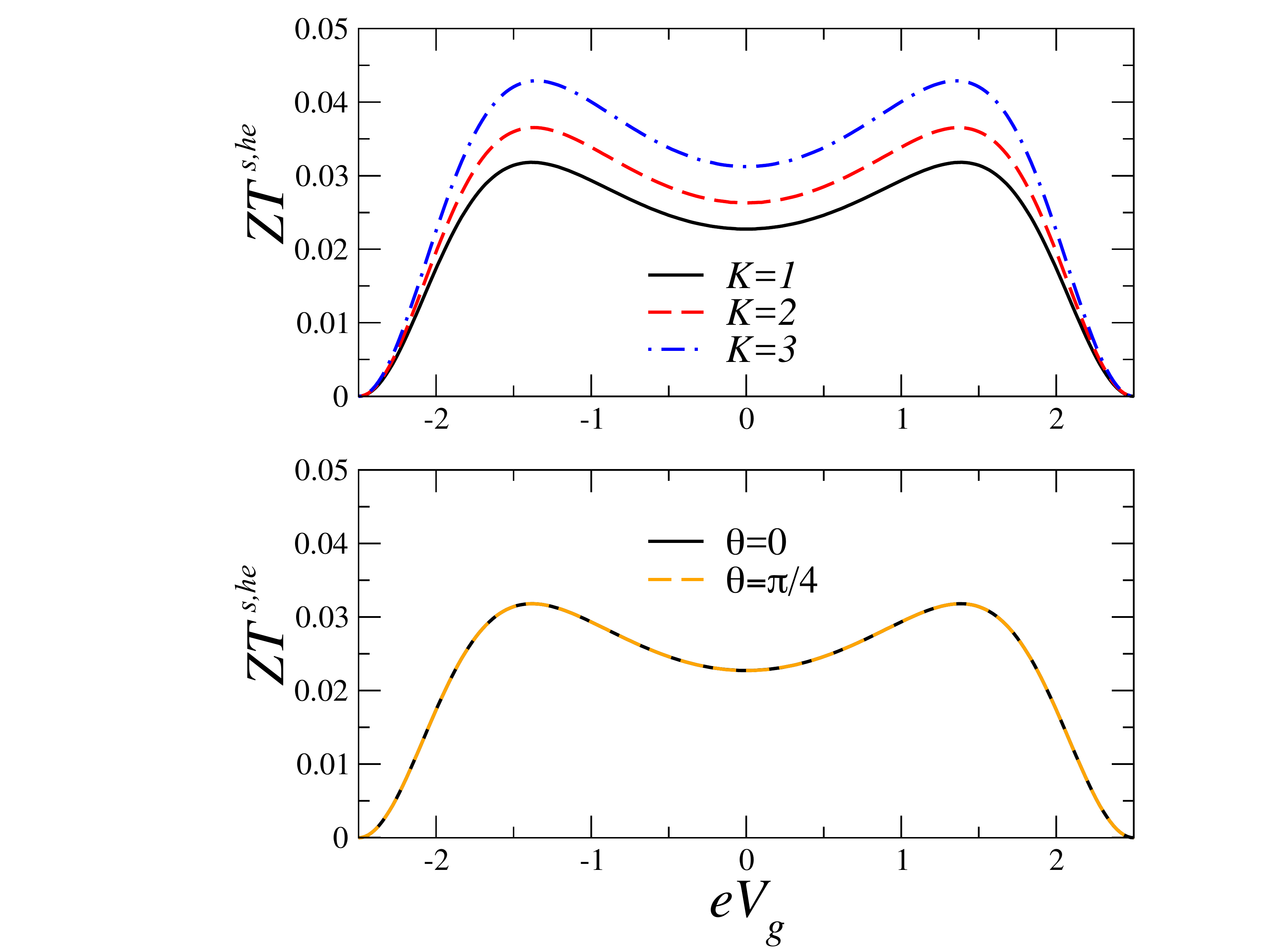}
\end{center}
\caption{(Color online) Figure of merit $ZT^{s,he}$  
corresponding at the spin heat engine. Other details are the same as in Fig. \ref{regimen-2a}. Other details are the same as in the previous Fig.}
\label{regimen-2b}
\end{figure}

In the lower panels of figure (\ref{regimen-2a}), we show the behavior of these coefficients when the direction of the magnetic field is tilted in an angle $\theta$ with respect to the direction of the spin orbit of the QSH.
 In contrast to the charge regime, the factors $u^{2}_{s,\sigma}$ enter  the coefficients in combinations like $u^{2}_{s,\uparrow}-u^{2}_{s,\downarrow}=\pm(\cos^{2}(\theta/2)-\sin^{2}(\theta/2))$ leading to a dependence with $\theta$. 
Explicitly, within the low-$T$ limit,   we have for ${\cal L}^{s}_{11}$
\begin{eqnarray}\label{theta-dependent}
{\cal L}^{s}_{11}&=&L_{11}^{\uparrow}-L_{11}^{\downarrow}\simeq T \sum_{s,\sigma} ( u^{2}_{s,\uparrow}-u^{2}_{s,\downarrow}) \rho_{d, s }(0) I_0 (\overline{K})\nonumber\\
                 &=&T (\cos^{2}(\theta/2)-\sin^{2}(\theta/2))\sum_{s} \rho_{d, s }(0) I_0 (\overline{K}).
\end{eqnarray}
A similar expression is found for ${\cal L}^{s}_{12}$.
An interference between the two spin  channels is clearly seen as a function of $\theta$.  In particular, when the field is perpendicular  to the direction of the spin-orbit of  the bar, the two coefficients vanish.

The efficiency of the device in this regime is analyzed in figure (\ref{regimen-2b}), where the figure of merit $(ZT)^{s,he}$ defined in Eq. (\ref{chi}), is shown 
 for the same parameters of Fig. \ref{regimen-2a}. For this mode, the performance is also enhanced for interacting systems with $K \neq 1$.
Interestingly, the effect of the alignment of the magnetic field affects identically ${\cal L}^{s}_{11}$ and ${\cal L}^{s}_{12}$. Hence, the vale of $ZT$ is not affected by the tilt angle $\theta$ of the magnetic field, as can be seen in the bottom panel of Fig.
(\ref{regimen-2b}).

\subsection{Spin refrigerator}
Finally, we discuss the spin refrigerator mode introduced in Section \ref{oper-3}. As in the heat engine, this regime operates only at finite magnetic field $B \neq 0$.

The coefficient of performance of the device in this case can be quantified by the ratio between the power developed by the polarized heat  flux flowing along the helical edge states $J^{Q,s}$ and the charge
flux $J^C$, between the external reservoir and the TI bar, as defined in Eq. (\ref{eta-s-fr}).  Following the same procedure as in the case of the spin
 heat engine, we  fix $\Delta T$ and maximize Eq. (\ref{eta-s-fr}). 
 The maximum value is 
\begin{equation} \label{maxsfr}
\eta^{s,fri,{\rm max}}= \eta_c \frac{ \chi^{fr} \sqrt{1 + (ZT)^{s,fr}} - 1}{\sqrt{1 + (ZT)^{s,fr}} + 1}.
\end{equation}
We have used the following definitions
\begin{equation}\label{chif}
\chi^{fr}= \frac{{\cal L}_{21}^s}{ {\cal L}_{12}^c}, \;\;\;\;\;\;\;\;\;\;\;\; (ZT)^{s,fr}  = \frac{{\cal L}_{21}^s {\cal L}_{12}^c }{ \mbox{Det}\widehat{\cal L}^{s,fr}},
\end{equation}
being $\mbox{Det}\widehat{\cal L}^{s,fr}$  the determinant of the matrix of Eq. (\ref{spinq}). The voltage bias corresponding to the 
maximum is
\begin{equation}\label{tdmuf}
 T \Delta \mu = - \Delta T \; \frac{{\cal L}_{22}^s}{{\cal L}_{21}^s} \left( 1 - \sqrt{1-R^{fr}} \right),
\end{equation} 
where $R^{fr}=\frac{{\cal L}_{12}^c {\cal L}_{21}^s}{ {\cal L}_{11}^c{\cal L}_{22}^s}$.
In this mode the maximum coefficient of performance at fixed $\Delta T$ is also  parametrized by a figure of merit $(ZT)^{s,fr}$, which has the same formal expression as for the modes previously analyzed.
 As in the other cases, the maximum is achieved for 
$(ZT)^{s,fr} \rightarrow \infty $. It is bounded by $\chi^{fr}  \, \eta_c$ and $\chi^{fr} \leq 1$,  with the maximum corresponding to a fully polarized current. In addition to the definitions of Eqs. (\ref{maxsfr}), (\ref{chif}) and
(\ref{tdmuf}), the following conditions must be satisfied
\begin{eqnarray}\label{sfr}
& &  T \Delta \mu <-\frac{{\cal L}^c_{22}}{ {\cal L}^s_{21}} \Delta T,\;\;\;\;\;\;\;\;\;\;\;\;\;{\cal L}^s_{21} >0,\nonumber \\
& &  T \Delta \mu >  -\frac{{\cal L}^c_{22}}{ {\cal L}^s_{21}}  \Delta T ,\;\;\;\;\;\;\;\;\;\;\;\;\;{\cal L}^s_{21} <0.
\end{eqnarray}

\begin{figure}[tbp]
\vspace{1.cm}
\begin{center}
\includegraphics[width=\columnwidth]{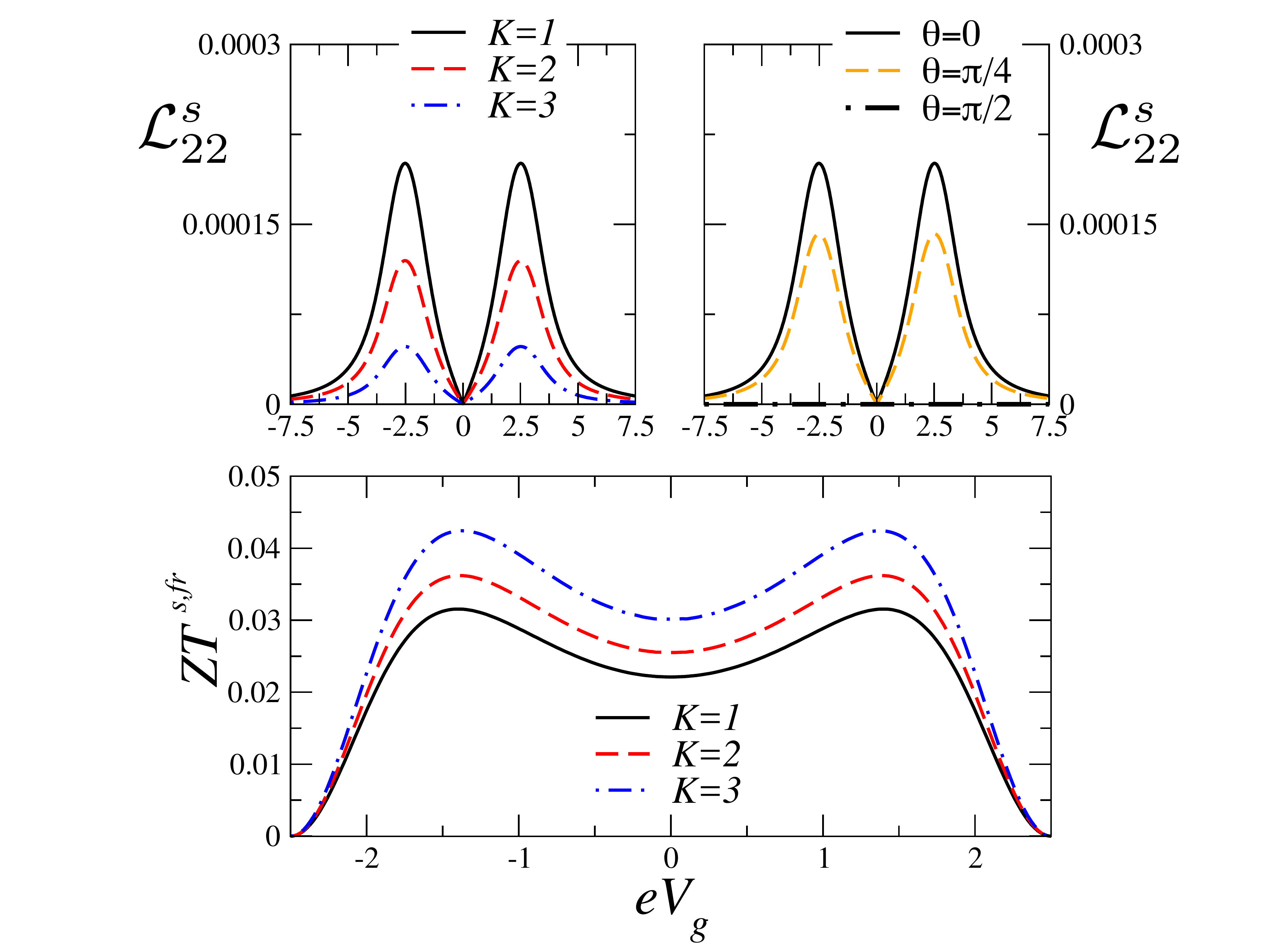}
\end{center}
\caption{(Color online) Onsager coefficients ${\cal L}^s_{22}$ (top panels) and figure of merit $ZT^{s,fr}$ (lower panel)  
corresponding to the spin refrigerator. Other details are the same as in Fig. \ref{regimen-2a}. Other details are the same as in the previous Fig. }
\label{regimen-3}
\end{figure}

In this case, the regime is characterized by the coefficients ${\cal L}^{s}_{22}$, plotted in the top panels of figure (\ref{regimen-3}), and ${\cal L}^{s}_{21}$. 
As in the spin heat operational mode, this regime implies a proper definition of in the sign $\xi$, such that ${\cal L}^{s}_{22}=\xi(L_{22}^{\uparrow}-L_{22}^{\downarrow})\ge0$ to characterize the thermoelectric response.
The dependence of the coefficients with the gate voltage and magnetic field is similar to that observed for the spin heat engine.  We have verified that the sign $\xi$, as a function of the gate voltages and magnetic field, is exactly the same within the spin heat and refrigerator regimes. This can be easily verified from  the low-temperature expressions of the coefficients in Eq. (\ref{coeff-approx}), by noticing that the only difference between $L_{11}^{\sigma}$ and $L_{22}^{\sigma}$ is given by the amplitude introduced by the factor $I_n(\overline{K})$ in Eq. (\ref{integrales}). As a consequence, the coefficient ${\cal L}^{s}_{21}$ is equal to ${\cal L}^{s}_{12}$ shown in Figure (\ref{regimen-2a}).
The coefficient ${\cal L}^{s}_{22}$ and the  figure of merit are shown in Fig. \ref{regimen-3}.

\section{Summary and discussion}
We have analyzed the thermoelectric response of a pair of helical edge states of a topological insulator coupled to an extra reservoir through a side quantum dot. By applying a gate voltage as well as a magnetic field at the quantum dot, 
different thermoelectric operational modes can be induced in this device. These include usual charge heat engine and refrigerator, as well as spin heat engine and refrigerator. The latter modes imply the conversion between heat and spin currents and are of interest in
spintronics. 

In this work we introduced a linear response description, relevant for small bias voltages and temperature differences and characterize the different regimes by efficiencies or coefficients of performance parametrized by figures of merit.
Considering a typical value for the parameter $\gamma \sim 1 \mu eV$,\cite{enhan} the spin heat engine and spin refrigerator regimes can be achieved with magnetic fields $B \sim 0.1 T$, which are of the order of magnitude of 
the ones implemented in laboratory to study quantum dots in the Kondo regime. \cite{kon-dot} Given the additional fact that our setup is based on a quantum dot coupled albeit not hosted by the topological insulator in the QSH state,
the application of a magnetic field is not expected to affect the nature of the helical edge states. For these estimates, the temperatures indicated in Fig. 6 range from $T=5 mK$ to $T=40 mK$ and we see that values of figure of merit
as large as $ZT=5$ can be achieved in the case of non-interacting helical edges. 
As in the case of the fractional quantum Hall effect studied in Ref. \onlinecite{enhan}, the thermoelectric performance is improved when the many-body interactions are relevant.
 Hence , for the cases with $K\neq 1$, the above estimates can be improved in a factor depending on $\overline{K}$, as shown in Eq. (\ref{ztlt}), and Figs 4,7,8 and 9.

\section*{Acknowledgments}
LA thanks the Kavli Institute for Theoretical Physics (KITP) for its hospitality during the program Thermodynamics of quantum systems: Measurement, engines and control.  EF thanks the Department of Physics of the the Faculty of Exact and Natural Sciences of the University of Buenos Aires for its hospitality.
We acknowledge support from CONICET and  MINCyT, Argentina, as well as  the Alexander von Humboldt Foundation, Germany. This work as supported in part by the National Science Foundation under Grant No. NSF PHY-1748958 at the KITP (LA), and under Grant No. DMR-1725401 at the University of Illinois (E.F.).

\appendix

\section{Density of states of the helical edges}\label{green}
The derivation of the density of states of the chiral Luttinger liquid $ \rho_{\sigma}(\omega)$ entering in Eq. (\ref{rho-lutt}) follows from the Hamiltonian in Eq. (\ref{Hll}) and can be obtained by using standard bosonization technique as presented in Ref. (\onlinecite{gia}). Here we summarize the main steps. 

Starting from the definition of the greater Green function $ig^{>}_{\downarrow}(t-t')$ and the expression of the fermionic fields in terms of the bosonic ones in Eq. (\ref{fermion-fields}) one arrives at
\begin{eqnarray}\label{gdn>}
 ig^{>}_{\downarrow}(t-t')&=&\big< \psi_{\downarrow}(t)\psi^{\dagger}_{\downarrow}(t')  \big>\nonumber\\
                        &=&\frac{1}{2\pi a}\big< e^{ iK_{+}\phi_{\downarrow}(t)} e^{ iK_{-}\phi_{\uparrow}(t)}
                        e^{-iK_{+}\phi_{\downarrow}(t')}e^{-iK_{-}\phi_{\uparrow}(t')} \big>\nonumber\\
                        &=&\frac{1}{2\pi a}e^{K^{2}_{+}D^{>}_{\downarrow}(t-t')}e^{K^{2}_{-}D^{>}_{\uparrow}(t-t')},
\end{eqnarray}
where  $D^{>}_{\sigma}(t-t')=D_{\sigma}^{<}(t'-t) =< \phi_{\sigma}(t)\phi_{\sigma}(t') >-\frac{1}{2}<\phi^2_{\sigma}(t) > -\frac{1}{2}<\phi^2_{\sigma}(t')>$, for $t>t'$ is the bosonic propagator associated at the $\sigma$ channel in Eq. (\ref{Hll}). The latter is given by the following expression
                                                   
\begin{eqnarray}\label{boson-correlation-lesser}
D_{\sigma}^{>}(t-t')&=& K\mbox{ln}\Bigl[ \frac{\mbox{sinh}(ia\pi T) }
              {\mbox{sinh}\bigl[ \pi T(t'-t+i a) \bigr]} \Bigr]. 
\end{eqnarray}

Inserting Eq. (\ref{boson-correlation-lesser}) into Eq. (\ref{gdn>}) the greater Green function can be simple written as 
\begin{eqnarray}\label{gdn>2}
 g^{>}_{\downarrow}(t-t')&=&\frac{-i}{2\pi a}\Bigl[\frac{\mbox{sinh}(ia\pi T)}{\mbox{sinh}\bigl[\pi T(t'-t+ia)\bigr]}\Bigr]^{ ( K^{2}_{+}+K^{2}_{-})K}\nonumber\\
 g^{>}_{\downarrow}(t-t')&=&\frac{-i}{2\pi a}\Bigl[\frac{\mbox{sinh}(ia\pi T)}{\mbox{sinh}\bigl[\pi T(t'-t+ia)\bigr]}\Bigr]^{\bar{K}},
\end{eqnarray}
where  $\bar{K}\equiv( K^{2}_{+}+K^{2}_{-})K=\frac{1}{2}(\frac{1}{K}+K)$.

The lesser Green function is obtained through the relation $g^{<}_{\sigma}(t-t')=-g^{>}_{\sigma}(t'-t)$,
\begin{eqnarray}\label{gdn<}
 g^{<}_{\downarrow}(t-t')&=&\frac{i}{2\pi a}\Bigl[\frac{\mbox{sinh}(ia\pi T)}{\mbox{sinh}\bigl[\pi T(t-t'+ia)\bigr]}\Bigr]^{\bar{K}},
\end{eqnarray}
Note that the left and right movers share the same temperature and, as a consequence, there is no spin dependence in the explicit expressions of the above Green functions.

Transforming Fourier to the frequency domain we have                                                   
\begin{eqnarray}
 g^{<}_{\sigma}(\omega)&=& i a^{\bar{K}-1} \frac{(2\pi T)^{\bar{K}-1}}{2\pi \Gamma(\bar{K})} e^{-\omega/2T}
              \Big\vert \Gamma(\bar{K}/2 + i\omega/2\pi T) \Big\vert^2.\nonumber\\
\end{eqnarray}
Finally, the density of states $ \rho_{\sigma}(\omega)$ can be introduced through the identity $g^{<}_{\sigma}(\omega)=2\pi i \rho_{\sigma}(\omega)f(\omega)$, where $f(\omega)$ is the Fermi distribution and $\Gamma(z)$ is the Gamma function.
The final expression for the densities of states, Eq. (\ref{rho-lutt}) reads 
\begin{eqnarray}
 \rho_{\sigma}(\omega)&=&a^{\bar{K}-1} \frac{(2\pi T)^{\bar{K}-1}}{\Gamma(\bar{K})} 
              \Big\vert \frac{\Gamma(\bar{K}/2 + i\omega/2\pi T)}{\Gamma(1/2 + i\omega/2\pi T)}\Big\vert^2.
\end{eqnarray}


\begin{thebibliography}{9}
\bibitem{wen}X. G. Wen, Topological orders and Edge excitations in FQH
states, Adv. Phys. {\bf 44}, 405 (1995).

\bibitem{halperin}B. I. Halperin, Quantized Hall conductance, current-carrying
edge states, and the existence of extended states in a two-
dimensional disordered potential, Phys. Rev. B {\bf 25}, 2185 (1982).

\bibitem{laughlin}R. B. Laughlin, Quantized Hall conductivity in two dimensions,
Phys. Rev. B {\bf 23}, 5632 (1981).

\bibitem{buttiker}M. B\"uttiker, Absence of backscattering in the quantum Hall effect 
in multiprobe conductors, Phys. Rev. B 38, 9375 (1988).

\bibitem{fradkin} E. Fradkin, {\sl Field Theories of Condensed Matter Systems}, 2nd
  edition, Cambridge University Press, (Cambridge, UK, 2013).
  
  \bibitem{ti1}C. L. Kane and E. J. Mele, $Z_2$
  Topological Order and the Quantum Spin Hall Effect, Phys. Rev. Lett. {\bf 95}, 146802
(2005).

\bibitem{ti2}C. L. Kane and E. J. Mele, Quantum Spin Hall Effect in Graphene, Phys. Rev. Lett. {\bf 95}, 226801 (2005).

\bibitem{ti3}B. A. Bernevig, T. L. Hughes, and S.-C. Zhang, Quantum Spin Hall Effect and Topological Phase Transition in HgTe Quantum Wells, Science {\bf 314}, 1757 (2006).

\bibitem{ti5} C. Wu, B. A. Bernevig, and S.-C. Zhang, Helical Liquid and the Edge of Quantum Spin Hall Systems, Phys. Rev. Lett. {\bf 96}, 106401 (2006).

\bibitem{ti6}C. Xu and J. Moore, Stability of the quantum spin Hall effect: Effects of interactions, disorder, and 
$Z_2$
 topology Phys. Rev. B {\bf 73}, 045322 (2006).


\bibitem{ti7} M. K\"onig, S. Wiedmann, C. Br\"une, A. Roth, H. Buhmann,
L. W. Molenkamp, X. L. Qi, and S.-C. Zhang, Quantum Spin Hall Insulator State in HgTe Quantum Wells, Science {\bf 318},
766 (2007).



\bibitem{k-f-1}  C. L. Kane, and M. P. A. Fisher,  Thermal Transport in a Luttinger Liquid,
Phys. Rev. Lett {\bf 76}, 3192 (1996). 
  
\bibitem{k-f-2} C. L. Kane, and M. P. A. Fisher, 
Quantized thermal transport in the fractional quantum Hall effect, 
Phys. Rev. B {\bf 55}, 15832 (1997). 


\bibitem{chamon}C. de C. Chamon and E. Fradkin, How to observe distinct universal conductances in tunneling to quantum Hall states: having the right contacts, Phys. Rev. B {\bf 56}, 2012 (1997).

\bibitem{eun-ah}E-A. Kim and E. Fradkin, Interedge tunneling in quantum Hall line junctions, Phys. Rev. B {\bf 67}, 045317 (2003)

\bibitem{chang}A. M. Chang, Chiral Luttinger liquids at the fractional quantum
Hall edge, Rev. Mod. Phys. {\bf 75}, 1449 (2003).

\bibitem{glattli}C. Glattli, Tunneling Experiments in
the Fractional Quantum Hall Effect Regime, S\'minaire Poincar\' {\bf 2}, 75 (2004)

 \bibitem{rosenow} S. Takei, M. Milletarì, and B. Rosenow, Phys. Rev. B 82,
041306 (2010).

\bibitem{torsten}T. Karzig, G. Refael, L. I. Glazman, F. von Oppen,
Energy Partitioning of Tunneling Currents into Luttinger Liquids,
 Phys. Rev. Lett. {\bf 107}, 176403 (2011).
 
 \bibitem{enhan} 
P. Roura-Bas, L. Arrachea, E. Fradkin, Enhanced thermoelectric response in the fractional quantum Hall effect
 Phys. Rev. B {\bf 97}, 081104 (2018)

\bibitem{gia}T. Giamarchi, Quantum Physics in One Dimension, Oxford University Press, Oxford (2004).
  
  
  \bibitem{granger}  J. P. Eisenstein and J. L. Reno, Observation of Chiral Heat Transport in the Quantum Hall Regime, G. Granger,
  Phys. Rev. Lett. {\bf 102}, 086803 (2009).



\bibitem{us}  L. Arrachea and E. Fradkin, 
Chiral heat transport in driven quantum Hall and quantum spin Hall edge states,
Phys. Rev. B {\bf 84}, 235436
  (2011).
  
\bibitem{Nam} 
S. G. Nam,  E. H. Hwang   and  H. J. Lee,Thermoelectric detection of chiral heat transport in graphene in the quantum Hall regime,
Phys.   Rev.   Lett. {\bf 110} 22680 (2013).

\bibitem{qcond} 
S. Jezouin,  F D.  Parmentier, A.   Anthore, U.   Gennser, A. J.    Cavanna   and F.  Pierre,   Quantum limit of heat flow across a single electronic channel,  
Science {\bf 342}
601  (2013).


%%%%%%%%%%
\bibitem{cappelli}  A. Cappelli, M. Huerta, and G. Zemba, 
Thermal Transport in Chiral Conformal Theories and Hierarchical Quantum Hall States, Nucl. Phys. B
  \textbf{636}, 568 (2002).
%%%%%%%%%%%


%%%%%%%%%%
\bibitem{grosfeld} 
E. Grosfeld and S. Das, 
Probing the Neutral Edge Modes in Transport across a Point Contact via Thermal Effects in the Read-Rezayi Non-Abelian Quantum Hall States,
Phys. Rev. Lett.
  \textbf{102}, 106403 (2009).
%%%%%%%%%%




  
  
  
 \bibitem{us1} H. Aita, L. Arrachea, C. Na\'on, and E. Fradkin,
 Heat transport through quantum Hall edge states: Tunneling versus capacitive coupling to reservoirs,
Phys. Rev. B {\bf 88}, 085122 (2013).


  



\bibitem{stern} G. Viola, S. Das, E. Grosfeld, and A. Stern, Thermoelectric probe for neutral edge modes in the fractional quantum Hall regime,
  Phys. Rev. Lett. {\bf 109}, 146801 (2012).
  

\bibitem{heiblum} I. Gurman, R. Sabo, M. Heiblum, V. Umansky, and
  D. Mahalu, Extracting net current from an upstream neutral mode in the fractional quantum Hall regime,
  Nature Comm. {\bf 3}, 1289 (2012).


\bibitem{altimiras} C. Altimiras, H. le Sueur, U. Gennser, A. Cavanna,
  D. Mailly and F. Pierre, Tuning energy relaxation along quantum Hall channels, 
  Phys. Rev. Lett. \textbf{105}, 226804
  (2010).

\bibitem{yacoby} 
V. Venkatachalam, S. Hart, L. Pfeiffer, K. West, and
  A. Yacoby, Local thermometry of neutral modes on the quantum Hall edge
  Nature Physics {\bf 8}, 676 (2012).

\bibitem{altimiras2}
  A. Cavanna, D. Mailly, and F. Pierre, 
  Chargeless Heat Transport in the Fractional Quantum Hall Regime, C. Altimiras, H. le Sueur, U. Gennser, A. Anthore,
  Phys. Rev. Lett. {\bf 109},
  026803 (2012).



  

\bibitem{baner} M. Banerjee, M. Heiblum, A. Rosenblatt, Y. Oreg, D. E. Feldman, A. Stern, and V. Umansky, 
Observed quantization of anyonic heat flow,
Nature {\bf 545}, 75 (2017).

 \bibitem{pheno}
Amit Aharon, Yuval Oreg, Ady Stern, Phenomenological theory of heat transport in the fractional quantum Hall effect,  arXiv:1805.09229



 \bibitem{rafa} R. S‡\'anchez, B. Sothmann, A. N. Jordan,  
Chiral thermoelectrics with quantum Hall edge states,
Phys. Rev. Lett. {\bf 114}, 146801 (2015).

\bibitem{peter}P. Samuelsson, S. Kheradsoud, B. Sothmann
Optimal quantum interference thermoelectric heat engine with edge states,
Phys. Rev. Lett. {\bf 118}, 256801 (2017) 

\bibitem{vanuci} 
L Vannucci, F Ronetti, G Dolcetto, M Carrega, M Sassetti, Interference-induced thermoelectric switching and heat rectification in quantum Hall junctions
Phys. Rev. B {\bf 92}, 075446 (2015).


\bibitem{jauho}A Xiao-Qin Yu, Zhen-Gang Zhu, Gang Su, A. -P. Jauho, Spincaloritronic battery, Phys. Rev. Applied {\bf 8}, 054038 (2017)



\bibitem{rone}Flavio Ronetti, Luca Vannucci, Giacomo Dolcetto, Matteo Carrega, Maura Sassetti,
Spin-thermoelectric transport induced by interactions and spin-flip processes in two dimensional topological insulators, 
 Phys. Rev. B {\bf 93}, 165414 (2016)
 
\bibitem{roda}  Sun-Yong Hwang, Rosa Lopez, Minchul Lee, David Sanchez, Nonlinear spin-thermoelectric transport in two-dimensional topological insulators,
 Phys. Rev. B {\bf 90}, 115301 (2014)

\bibitem{graph}Po-Hao Chang, Mohammad Saeed Bahramy, Naoto Nagaosa, Branislav K. Nikolic
Giant thermoelectric effect in graphene-based topological insulators with nanopores,
 Nano Lett. {\bf 14}, 3779 (2014) 

\bibitem{bjorn}D. G. Rothe, E. M. Hankiewicz, B. Trauzettel, M. Guigou,
Spin-dependent thermoelectric transport in HgTe/CdTe quantum wells,  Phys. Rev. B {\bf 86}, 165434 (2012) 

\bibitem{soth}
B. Sothmann and E. M. Hankiewicz, Fingerprint of topological Andreev bound states in phase-dependent heat transport,
Phys. Rev. B {\bf 94}, 081407(R) (2016)

\bibitem{benj} Arjun Mani, Colin Benjamin, Helical thermoelectrics and refrigeration, Phys. Rev. E {\bf 97}, 022114 (2018).

\bibitem{linke} M. Josefsson, A. Svilans, A. M. Burke, E. A. Hoffmann, S. Fahlvik, C. Thelander, M. Leijnse, H. Linke,
A quantum-dot heat engine operating close to the thermodynamic efficiency limits, arXiv:1710.00742


\bibitem{casati}G. Benenti, G. Casati, K. Saito, R. S. Whitney, Fundamental aspects of steady-state conversion of heat to work at the nanoscale,
Phys. Rep. {\bf 694}, 1 (2017).

\bibitem{ora}O.Entin-Wohlman,Y.Imry and A.Aharony, Enhanced performance of joint cooling and energy production, Phys.Rev. B {\bf 91} 054302 (2015).


\bibitem{thir}
H. Thierschmann, R. S\'anchez, B. Sothmann, F. Arnold, C. Heyn, W. Hansen, H. Buhmann, and L, W. Molenkamp, Three-terminal energy harvester with coupled quantum dots,
Nature Nanotechnology {\bf 10}, 854 (2015)

 \bibitem{tang} G. Tang, J. Thingna, and J. Wang, Thermodynamics of energy, charge, and spin currents in a thermoelectric quantum-dot spin valve,
Phys. Rev B {\bf 97}, 155430 (2018)

\bibitem{dolcini}F. Dolcini, Full electrical control of charge and spin conductance through interferometry of edge states in topological insulators, Phys. Rev. B {\bf 83}, (2011).

\bibitem{citro}F. Romeo, R. Citro, D. Ferraro and M. Sassetti, Electrical switching and interferometry of massive Dirac particles in topological insulator constrictions, Phys. Rev.
B {\bf 86}, 165418 (2012)


\bibitem{bruno1}
B. Rizzo, L. Arrachea, M. Moskalets, Transport phenomena in helical edge states interferometers. A Green's function approach, Phys Rev B {\bf 88}, 155433 (2013)

\bibitem{tidot1} F. Cr\'epin, J. C. Budich, F. Dolcini, P. Recher, and B.
Trauzettel, Renormalization group approach for the scattering off a single Rashba impurity in a helical liquid, Phys. Rev. B {\bf 86}, 121106 (R ) (2012).
\bibitem{tidot2} G. Dolcetto, F. Cavaliere, D. Ferraro, and M. Sassetti, Generating and controlling spin-polarized currents induced by a quantum spin Hall antidot, Phys. Rev. B {\bf 87}, 085425 (2013).
\bibitem{tidot3} J. V\"ayrynen, M. Goldstein, and L. I. Glazman, Helical Edge Resistance Introduced by Charge Puddles, Phys. Rev. Lett. {\bf 110}, 216402 (2013).
\bibitem{tidot4}J. V\"ayrynen, M. Goldstein, Y. Gefen, and L. I. Glazman, Resistance of helical edges formed in a semiconductor heterostructure,
Phys. Rev. B {\bf 90}, 115309 (2014).
\bibitem{tidot5}A. M. Lunde and G. Platero, Helical edge states coupled to a spin bath: Current-induced magnetization, Phys. Rev. B {\bf 86}, 035112 (2012).
\bibitem{tidot6} K. T. Law, C. Y. Seng, Patrick A. Lee, and T. K. Ng, Quantum dot in a two-dimensional topological insulator: The two-channel Kondo fixed point
Phys. Rev. B {\bf 81}, 041305 (2010).
\bibitem{tidot7}T. L. Schmidt, S. Rachel, F. von Oppen, and L. I. Glazman, Inelastic Electron Backscattering in a Generic Helical Edge Channel, Phys. Rev. Lett. {\bf 108}, 156402 (2012).
\bibitem{tidot8}V. Cheianov and L. I. Glazman, Mesoscopic Fluctuations of Conductance of a Helical Edge Contaminated by Magnetic Impurities, Phys. Rev. Lett. {\bf 110}, 206803 (2013).
\bibitem{tidot9}T. Posske, C-X Liu, J. C. Budich, and B. Trauzettel, Exact Results for the Kondo Screening Cloud of Two Helical Liquids, Phys. Rev. Lett. {\bf 110}, 016602 (2013).
\bibitem{tidot10}T. Posske and B. Trauzettel, Direct proportionality between the Kondo cloud and current cross correlations in helical liquids, Phys. Rev. B {\bf 89}, 075108 (2014).
\bibitem{tidot11} B. Probst, P. Virtanen, and P. Recher, Controlling spin polarization of a quantum dot via a helical edge state, Phys. Rev. B {\bf 92}, 045430 (2015).
\bibitem{tidot12}A. Rod, G. Dolcetto, S. Rachel and T. L. Schmidt, Transport through a quantum spin Hall antidot as a spectroscopic probe of spin textures, Phys. Rev. B {\bf 94}, 035428 (2016)
\bibitem{tidot13}
G. Tang,  F. Xu,  S. Mi, J. Wang, Spin-resolved electron waiting times in a quantum-dot spin valve Phys Rev B  {\bf 97}, 165407 (2018)





\bibitem{bruno2}B. Rizzo, A. Camjayi, Liliana Arrachea
Transport in quantum spin Hall edges in contact to a quantum dot
 Phys. Rev. B {\bf 94}, 125425 (2016)

 
 \bibitem{magdot1} Q. Meng, S. Vishveshwara, T. L. Hughes, Spin-transfer torque and electric current in helical edge states in quantum spin Hall devices,  Phys. Rev. B {\bf 90}, 205403 (2014)
 
 \bibitem{magdot2} L. Arrachea and F. von Oppen, Nanomagnet coupled to quantum spin Hall edge: An adiabatic quantum motor, Physica E {\bf 74}, 596 (2015)
 
 \bibitem{magdot3} 	
P. G. Silvestrov, P.  Recher, P. W.  Brouwer, Noiseless manipulation of helical edge state transport by a quantum magnet, Phys. Rev. B {\bf 93}, 205130 (2016)


\bibitem{kane-fish-92}
C. L. Kane and Matthew P. A. Fisher, Transmission through barriers and resonant tunneling in an interacting one-dimensional electron gas
Phys. Rev. B {\bf 46}, 15233 (1992)

\bibitem{kon-dot} S. M. Cronenwett, T. H. Oosterkamp, L. P. Kouwenhoven, 
A Tunable Kondo Effect in Quantum Dots, 
Science  {\bf 281}. 540 (1998)





 






\end{thebibliography}
\end{document}